\documentclass[final]{elsart}

\usepackage[dvips]{graphicx}
\usepackage{amssymb}
\usepackage{amsmath}
\usepackage{subfigure}
\usepackage{threeparttable}
\usepackage{url}

\usepackage{color}

\begin{document}

\begin{frontmatter}

\title{\Large{On the design of experiments based on plastic scintillators using {\sc Geant4} simulations}}

\author[ALCALA]{G. Ros\corauthref{cor1}},
\author[ALCALA]{G. S\'aez-Cano},
\author[UNAM]{G. A. Medina-Tanco},
\author[CONICET]{A. D. Supanitsky}.
\address[ALCALA]{Dpto. F\'isica y Matem\'aticas, Universidad de Alcal\'a. Alcal\'a de Henares (Spain).} 
\address[UNAM]{Instituto de Ciencias Nucleares, UNAM, M\'exico D. F. (Mexico).}
\address[CONICET]{Instituto de Astronom\'ia y F\'isica del Espacio, CONICET-UBA, Buenos Aires (Argentina).}
\corauth[cor1]{e-mail: german.ros@uah.es}
\begin{abstract}
Plastic scintillators are widely used as particle detectors in many fields, mainly, medicine, particle physics and astrophysics.
Traditionally, they are coupled to a photo-multplier (PMT) but now silicon photo-multipliers (SiPM) are evolving as a promising robust alternative, 
specially in space born experiments since plastic scintillators may be a light option for low Earth orbit missions. Therefore it 
is timely to make a new analysis of the optimal design for experiments based on plastic scintillators in realistic conditions in such a configuration.

We analyze here their response to an isotropic flux of electron and proton primaries in the energy range from 1 MeV to 1 GeV, 
a typical scenario for cosmic ray or space weather experiments, through detailed {\sc GEANT4} simulations. First, we focus on the effect of increasing 
the ratio between the plastic volume and the area of the photo-detector itself and, second, on the benefits of using a reflective coating around 
the plastic, the most common technique to increase light collection efficiency. In order to achieve a general approach,  it is necessary to 
consider several detector setups. Therefore, we have performed a full set of simulations 
using the highly tested {\sc GEANT4} simulation tool: several parameters have been analyzed such as the energy lost in the coating, the deposited energy 
in the scintillator, the optical absorption, the fraction of scintillation photons that are not detected, the light collection at the photo-detector,
the pulse shape and its time parameters and finally, other design parameters as the surface roughness, the coating reflectivity and the case of a 
scintillator with two decay components. This work could serve as a guide on the design of future experiments based on the use of plastic 
scintillators.

\end{abstract}


\end{frontmatter}

\section{Introduction}\label{sec:introduction}

Plastic scintillation detectors have been used in several fields for decades. As a tracker or calorimeter in nuclear and high energy physics 
thanks to their fast time response, high efficiency for charged particles, ease to manufacture, versatility and moderate costs. 
As an example, they have been recently selected for MINOS \cite{MINOS}, OPERA \cite{OPERA} and AugerPrime \cite{AugerPrime}, the extension of the 
Pierre Auger Observatory. A complete review on the use of scintillators in particle physics can be found in 
\cite{KharzheevReview}. Plastic scintillators can be exposed to high levels of radiation which along with their simplicity, low density and large volume compared 
to solid state based systems makes them also suitable for astrophysical purposes. Thus, plastic scintillators have been used as particle or 
neutron spectrometers in planetary missions in the past such as Phobos, Lunar Prospector or Mars Odyssey and, more recently, in Dawn and Solar 
Orbiter (for a review see \cite{Owens}), AMS \cite{AMS02} or DAMPE \cite{DAMPE}. In addition, their high detection efficiency and 
the proportionality between the light output and exciting energy make them very useful for radiotherapy and dossimetry applications \cite{Beddar}.

\begin{figure}
\centerline{
\subfigure{\includegraphics[width=8.5cm]{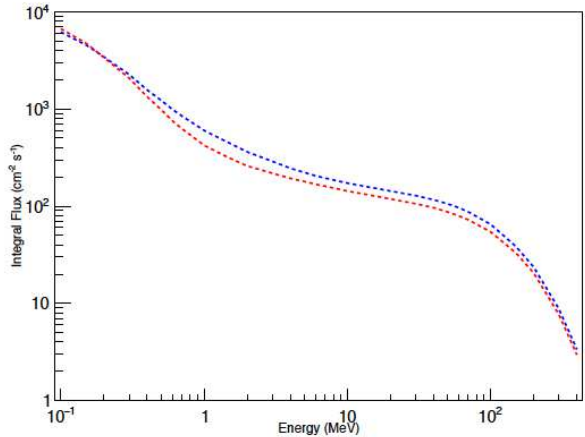}}
}
\caption{Integral flux of trapped protons in solar maximum (red) and minimum (blue) for a typical low earth orbit akin to the ISS 
(obtained using SPENVIS simulation code \cite{SPENVIS})}
\label{fig:AP8}
\end{figure}

In this study, we have selected as a source an isotropic flux of electrons and protons from 1 MeV to 1 GeV, a typical scenario in the detection 
of cosmic ray particles since there are three main sources of radiation at Low Earth Orbit (LEO) altitude, which are Solar Event Particles, Trapped 
protons and electrons and Galactic Cosmic Rays. As an example the integral flux of trapped protons at LEO is shown in Fig. \ref{fig:AP8}.
However, this selection is also of interest for any of the fields of application of plastic scintillators such as electron 
\cite{Hogstrom} or proton \cite{Newhauser} beam therapies or in case of hadronic calorimetres in particle physics as, for example, in 
CMS \cite{CMS}. 

We center on two issues of general interest, i.e., how, for a fixed photo-detector area, the increase of the plastic volume and the use of a 
reflective coating around the plastic could affect the capabilities of the experiment. The former is specially important in space experiments 
where the mass is a critical point. The latter is the most common technique to increase the light collection efficiency for a given volume,
the key point in most of these experiments.

In the literature, the effect of the scintillator volume has been previously analyzed in the medical area (dossimetry, radiotherapy) in case of plastics
using a beam of cobalt-60 radiation (electrons of 315 KeV and gammas of $\sim$1 MeV) \cite{Archambault} but also in liquid scintillators radiated 
with electrons and protons in the MeV range \cite{GALLOWAY} and with pure simulations \cite{Ghal-Eh}, but as far as we know it has not been analyzed 
in the astrophysical scenario of a space mission where the energy input covers an extended spectrum and the mass is a critical issue. 
The effect of the reflective coating has also been analyzed in \cite{MINOS, Dyshkant, Riggi} and a general discussion could be found in \cite{KharzheevReview}. 
However, most of the results are based on experimental setups with different geometries, scintillators, primary beams and photo-detectors. 
Therefore, the comparison between the results of these studies is not trivial. Alternatively, we have used the {\sc GEANT4} simulation tool
\cite{GEANT4} to consider a more general approach. {\sc GEANT4} allows to track photons inside the medium and to take into account all the 
optical properties of scintillators and its coating (emission, absorption, reflection, refraction, etc.). Furthermore, it has been demonstrated 
that {\sc GEANT4} realistically describes the response of plastic scintillation in space physics \cite{EspiritoSanto}, particle physics 
\cite{Kohley,Zhang} or dossimetry \cite{Guimaraes}.

\section{Simulation setup}\label{sec:Simulation}

\subsection{Geometry and material properties}\label{sec:GeomMaterial}

\begin{figure}
\centerline{
\subfigure{\includegraphics[width=8cm]{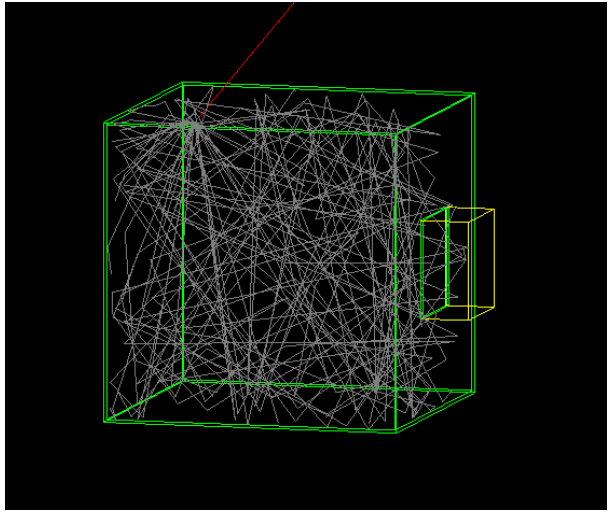}}
}
\caption{Setup for the simulation. The case for $F$=3 is shown. The green and yellow boxes represent the plastic scintillator and the 
photo-detector respectively, the red line is the incoming particle (1 MeV electron in this case) and the gray lines are the scintillation 
photons produced.}
\label{fig:Setup}
\end{figure}

We want to study the influence on light collection efficiency of the ratio between the volume of the plastic scintillator and the area of the 
photo-detector in contact with it. We assume the photo-detector as a silicon parallelepiped of 1 cm width and a contact area with the 
scintillator of $L \times L$, taking $L=2$ cm. It is located in one face of the plastic which is a cube whose volume is ${(F \times L)}^ 3$, 
where $F$ is the scale factor whose values will be $F=1, 3, 10$. The setup is shown in Fig. \ref{fig:Setup}. If a photon crosses the interface 
between the plastic scintillator and the photo-detector it is considered as {\it detected} and then its propagation is terminated.

Organic plastic scintillators have been widely produced using different techniques that lead to different plastic properties. These methods 
include the bulk polymerization method which achieves the highest transparency and uniformity but with high cost and manufacture time, 
injection molding that is highly productive (used at ATLAS and LHCb), the molding method (also used at CERN experiments), the extrusion 
method where plastic is produced using mechanically extruded polystyrene pellets (developed at Fermilab for MINOS \cite{MINOS}). A detailed 
description could be found in \cite{KharzheevReview}. In addition, a novel method based on the technology of photosensitivity rapid prototyping 
has been developed recently, which requires a shorter manufacture time, reduces the cost and allows to use the 3D printing technology \cite{Zhu16-1}.

\begin{figure}
\centerline{
\subfigure{\includegraphics[width=8.5cm]{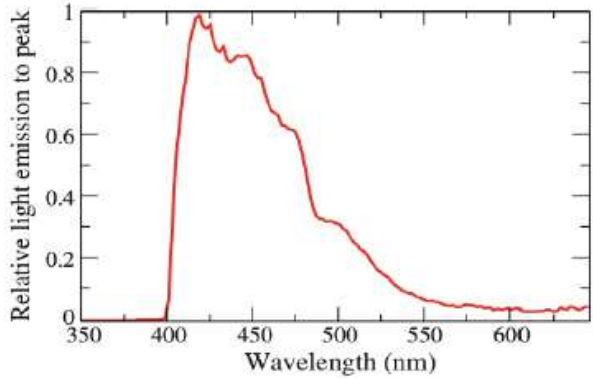}}
}
\caption{Emission spectrum of the doped plastic scintillator (taken from \cite{KikawaThesis}).}
\label{fig:plasticspectrum}
\end{figure}

We have selected the plastic developed at Fermilab for our simulation setup since the extrusion method has been recently improved for 
MINERVA \cite{MINERVA} and these plastics are widely used and tested in several cosmic ray experiments such the Pierre Auger Observatory 
\cite{AMIGA} and BATATA \cite{BATATA}. It is an extruded scintillator with a co-extruded reflective coating, whose manufacturing process has 
been optimized for higher light yield and lower costs. The base material in the plastic scintillates in the UV, 
but the mean free path of those photons is only a few millimeters, therefore, a wavelength shifter, also called 'fluor', needs to be added to 
the material. Thus, the scintillator is infused with two dopants, PPO and POPOP. The choice of the fluors is largely dictated by their emission 
and absorption spectra. Fluor absorption spectrum should be close to the base emission spectrum. The maxima of the wavelength shifter absorption 
and emission spectra should be as far away from each other (Stokes shift) as possible to avoid self absorption of emitted photons. The transmission 
achieved is more than 85\% in the spectral range of interest \cite{Plau-Dalmau99}. In addition, PPO and POPOP are a great choice 
for polystyrene-based plastic scintillators in order to achieve higher intensity and light yield if their concentration is properly tuned for 
each plastic \cite{Zhu16-2}. Thus, their concentration was optimized to be 1\% and 0.03\% by weight for PPO and POPOP respectively 
for the scintillator assumed here \cite{MINERVA}. The emission spectrum of the extruded plastic with these dopants is shown in 
Fig. \ref{fig:plasticspectrum}. The other properties of the plastic scintillator considered in the simulation are shown in Table \ref{table:plastic}.

\begin{table}
\caption{Plastic scintillator properties}
\label{table:plastic}
\begin{center}
\begin{tabular}{|c|c|c|c|c|c|} 
\hline
density & refraction 	& absorption 	& scintillation	& decay	& Birk's \\
        & index 	&length 	& yield		& time	& constant \cite{Torrisi} \\
\hline
 1.08 g/cm$^3$ &  1.58 	&  250 cm &	8000 ph./MeV & 3.6 ns &	0.126 mm/MeV \\
\hline
\end{tabular}
\end{center}
\end{table}

As previously mentioned, the scintillator is covered with a co-extruded reflective coating that is applied simultaneously with the extrusion of the 
scintillator. Thus, pellets of TiO$_2$ were mixed with polystyrene pellets. This method reduced significantly the costs and requires less 
manpower than wrapping the scintillator with a reflective material such as Tyvek, aluminum, Teflon or Mylar (see a comparative in \cite{Taheri}), ensuring at the same time that no spurious air layers are left in between the scintillator and the covering. 
The TiO$_2$ concentration and thickness of the coating to maximize the light output assuring uniformity during the manufacturing process has been
experimentally determined to be 15\% by weight and 0.25 mm respectively. However, the coating thickness is also important to limit the entrance 
of the primary particles into the plastic which determines the trigger rate in astrophysics experiments, so we have simulated several thicknesses 
of the coating (usually 0.25 and 0.50 mm) and a plastic scintillator without it. 

The entire assemble is surrounded by the "galactic vacuum" pre-defined in {\sc Geant4} (Z=1, A=1.01g/mole, $\rho$=1 $\times$10$^{-25}$ g/cm$^3$, 
P=3$\times$10$^{-18}$ Pa, T=2.73K).

\subsection{Primary flux and physical processes}\label{sec:Primary}

Our primary input consist of an isotropic flux of electrons and protons. The selected energies are 1, 10, 30, 100 MeV and 1 GeV. 
We have simulated 200 events per each primary particle type, energy, scale factor and for each coating thickness, which adds more 
than 18000 simulations.

All the expected physics processes and particles have been considered and activated in our {\sc Geant4} simulation. Standard electromagnetic 
processes, e.g. ionization, bremsstrahlung, multiple scattering, pair production, Compton scattering and photoelectric effect are considered.
The optical processes include scintillation, Cerenkov emission, bulk absorption, Rayleigh scattering and boundary processes 
(reflection, refraction, absorption). The dominant photon-generating process is scintillation where it is important to activate the 
Birk's effect. According to Birk's widely used model, the actual light yield of organic scintillators is reduced because 
of recombination and quenching effects between the excited molecules. This reduction is of $\sim$2\% 
\cite{DietzLaursonThesis}. In addition, special care must be taken with optical physics as it adds special numerical particles, called optical photons in the {\sc Geant4} parlance, 
that could be created in optical processes of our interest such as scintillation or Cerenkov radiation. Optical photons are the only particle 
that can be reflected or refracted at optical surfaces that will be described later in Section \ref{sec:Surfaces}.

\subsection{Characterization of interfaces}\label{sec:Surfaces}

The properties of the optical surfaces can be used to simulate a variety of surface conditions and are crucial for a realistic simulation.
By default, the optical surface processes are exclusively determined from the refractive indexes of the two materials forming the surface from which
the reflection or refraction angles (Snell's law) and probabilities (Fresnel equations) are calculated. However, this corresponds to the simulation of 
a perfectly smooth surface between two dielectrics. If other configurations have to be simulated, the surface properties have to be defined 
by the user. Several models and options are included in {\sc Geant4}. We have selected the {\it glisur} model for surface description which 
allows to include the degree of polishing of the surfaces. The roughness parameter could vary between 0 (completely diffuse, Lambertian reflection) 
and 1 (perfectly polished, reflection according to Fresnel's equations). We have selected 0.5 since the coating infusion is usually quite homogeneous 
but could not be considered as perfectly polished. An exhaustive description of the possibilities, the physics involved as well as technical issues 
can be found in \cite{DietzLaursonThesis}. 

If no coating is considered, the surface between plastic and vacuum should be defined. We select a {\it dielectric-dielectric} surface and the 
reflection and transmission probabilities are determined from Fresnel equations.

The correct simulation of the surface between plastic and coating is critical since the light collection depends not only on the scintillator 
transparency but also on the quality of the scintillator surface and reflective materials used. In \cite{MINOS} plastic strips with the co-extruded 
coating and with Tyvek wrapping are compared with no significant difference in light output, so that we have decided to simulate the 
plastic-coating interface with the optical properties of the latter. Tyvek is not a perfect diffuse reflector as Teflon but more similar 
to Aluminum. According to \cite{DietzLaursonThesis}, reflectivity has been set to 90\% in the whole wavelength range of interest, with 
15\% diffuse reflection and 85\% 'specular-lobe' reflection (based on micro-facet orientation which depends on the surface roughness parameter). 
This results in a predominant geometric reflection which is smeared by the surface roughness and a small diffuse fraction. In \cite{KharzheevReview} 
it is commented that the co-extruded coating achieves reflectivities around 96\% if TiO$_2$ concentration reaches 18\%. This is in agreement with 
the selected value since our coating has 15\% TiO$_2$ concentration so reflectivity is expected to be somewhat smaller. 

Since co-extruded coating behaves like aluminum or Tyvek reflectors, we have selected the surface type as {\it dielectric-metal} in the simulation.
Thus, optical photons arriving at the surface will be reflected with a probability according to the reflectivity value imposed while the rest will be
absorbed (no transmission). Strictly speaking, photons that are not reflected at such a surface will enter the next volume and undergo absorption 
corresponding to the attenuation length of the material, which is typically very short for optical photons in metals due to its large conductivity 
(for example it is around 22 $\mu$m in Aluminum in our frequency range \cite{SkinEffectCalculator}, much less than the thickness selected 
(150-500 $\mu$m)). The same approach has been used by other authors \cite{Riggi} and is usually recommended in Geant4 documentation. However, 
other authors as the ones of \cite{EspiritoSanto} prefer to use a {\it dielectric-dielectric} surface but with an additional option, that is to set the 
surface as \textquotedblleft groundfrontpainted\textquotedblright which, 
in fact, is equivalent since that fixes transmission to zero. As a consequence, the thickness of the coating will not affect the behavior
of optical photons produced inside the plastic volume so its effect is negligible when analyzing the optical absorption (Sec. \ref{sec:Absorption})
or the light collection efficiency (Sec. \ref{sec:Efficiency}) but it has a very important effect on the energy threshold of the experiment
(Sec. \ref{sec:Energy}).

\section{Energy Balance}\label{sec:Energy}

In this section we analyze the energy deposited by the primary flux in the scintillator and its transfer to scintillation photons. 
We select the base case with standard coating thickness of 0.25 mm unless otherwise is stated.

\subsection{Primaries impinging the photo-detector}

An issue that must be kept into consideration in the design of the experiment is that primaries could impact directly on the photo-detector.
In medical or accelerator physics where primary beams are used this will not happen but it is important in astroparticle experiments where primaries
arrive isotropically. From our simulation we get that the fraction of primaries impinging the photo-detector is around 25\%, 5\% and 2.0\% in case 
of $F$=1, 3 and 10 respectively, in overall agreement with the ratio between photo-detector and plastic areas. The detailed simulation 
of the photo-detector is a complex issue and beyond the scope of this work, so we just 
consider it as a silicon block. Within this limitation, we obtain that all the protons in the 1-30 MeV range deposit all their energy in the 
photo-detector and do not enter to the scintillator (they suffer several ionizations). On the other hand, 25\% and 50\% of 1 and 10 MeV electrons
traverse the photo-detector and reach the scintillator respectively, as do all the more energetic electrons. A full simulation of the 
assembly should be done in each experiment to fully understand the effects of this fact on their capabilities.

\subsection{Primaries that escape the assembly}

\begin{figure}
\centerline{
\subfigure{\includegraphics[width=9.5cm]{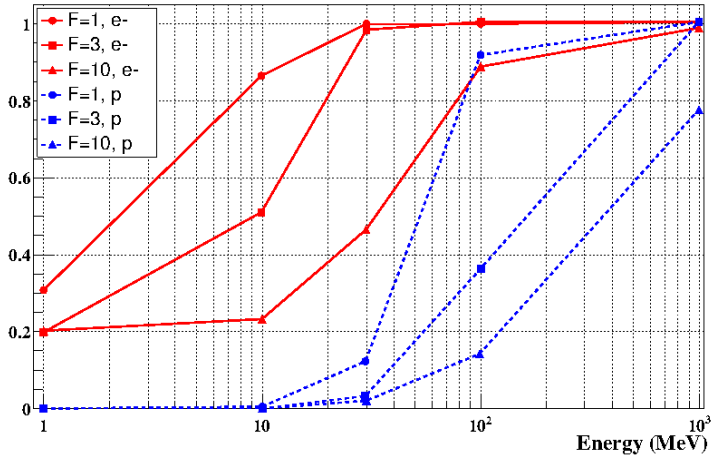}}
}
\caption{Fraction of primaries that do not deposit all their energy in the assembly.}
\label{fig:escape}
\end{figure}

Fig. \ref{fig:escape} shows the fraction of primaries that escape the detector without depositing all their energy in it.  As energy increases 
primary particles escape more easily from the detector. The volume of the plastic makes the largest difference at intermediate energies (10-100 MeV), 
while low energy particles are not able to escape (except for those that enter near the detector edges), 1 GeV particles do it almost
in any case. 

\subsection{Energy loss in the coating}

\begin{figure}
\centerline{
\subfigure{\includegraphics[width=8cm]{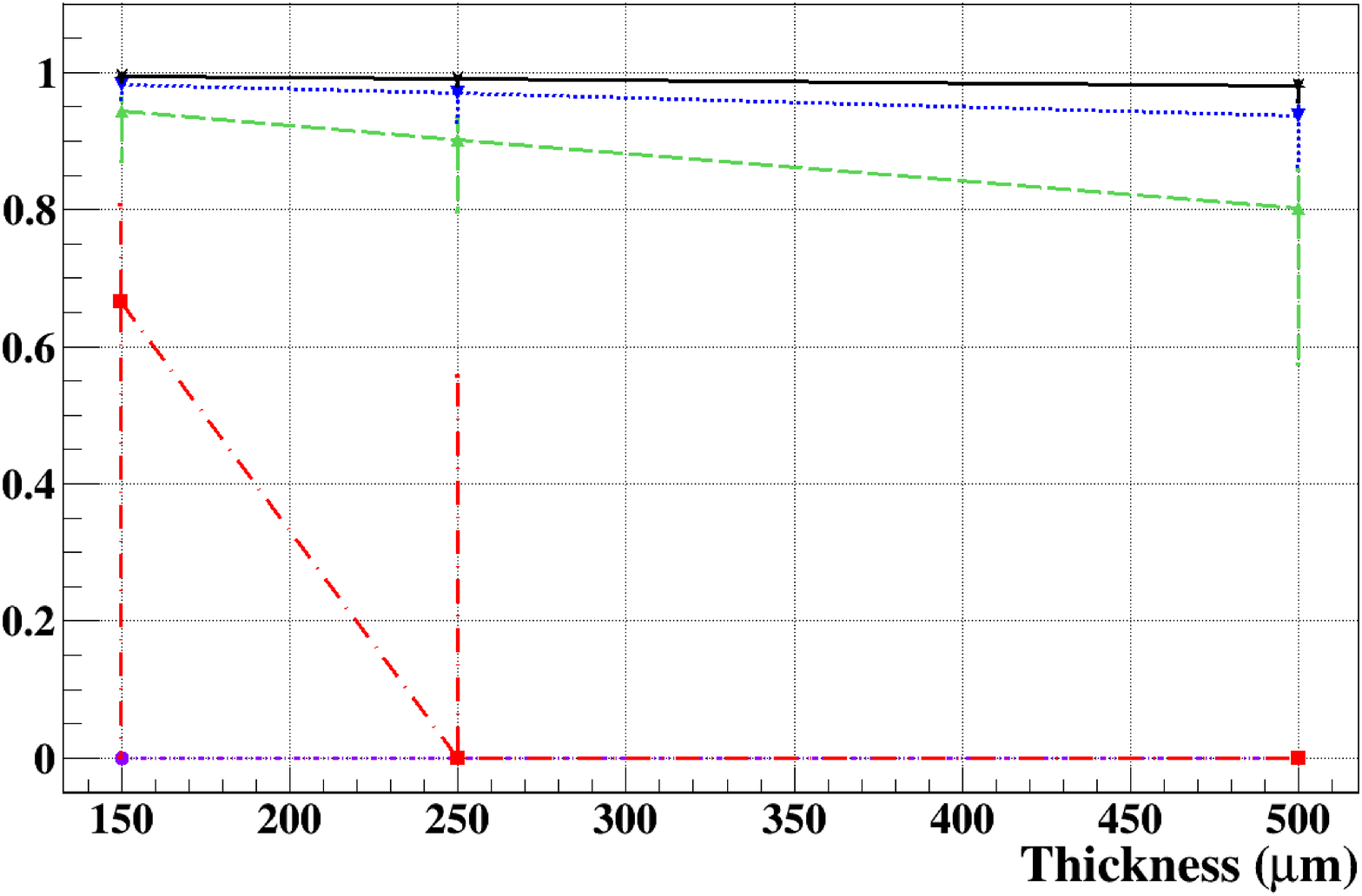}}
\subfigure{\includegraphics[width=8cm]{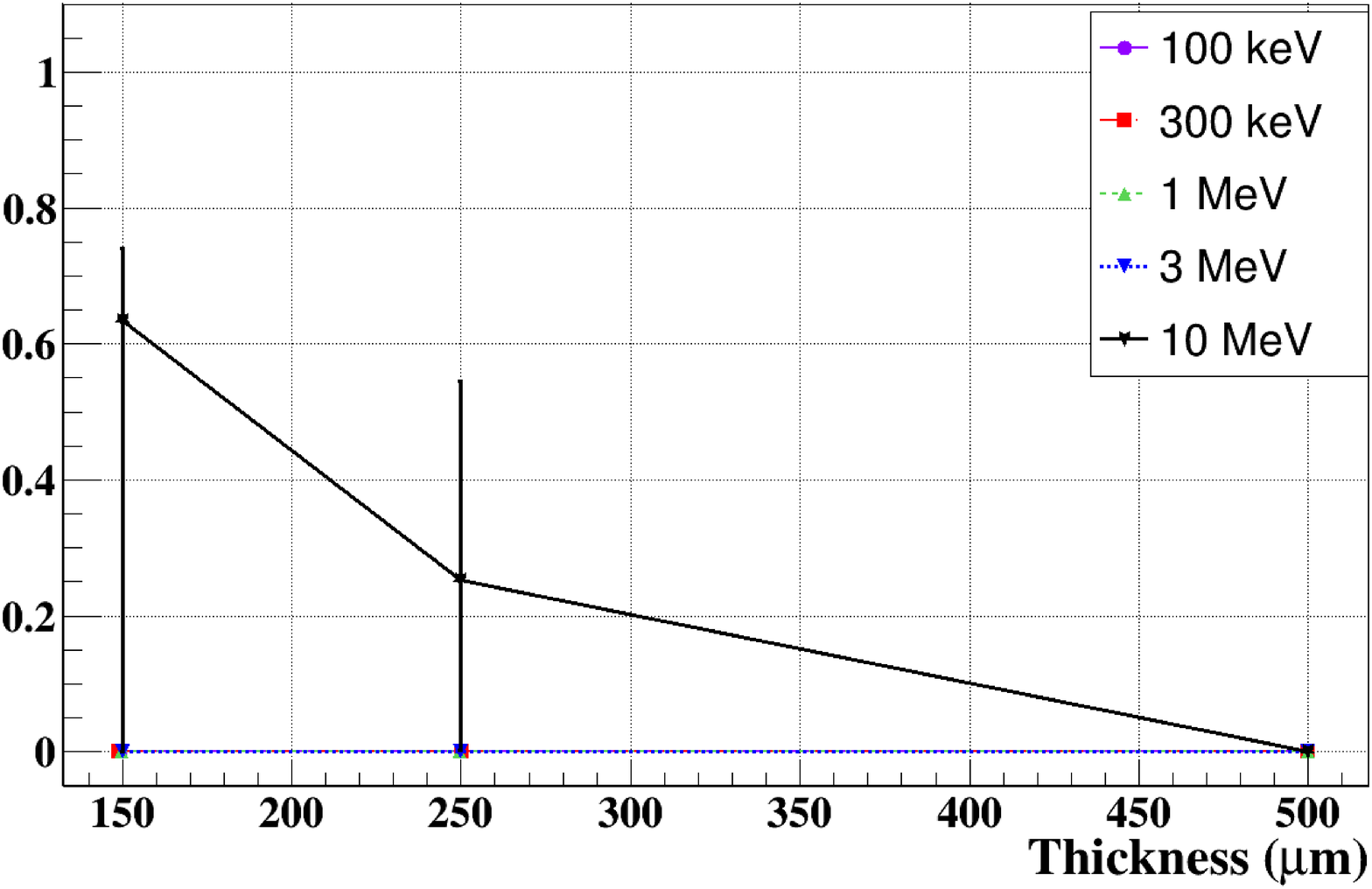}}
}
\caption{Fraction of the primary energy remaining after traversing the coating for electrons (left) and protons (right). Error bars
are the region of 68\% probability (in all the figures unless otherwise specified).}
\label{fig:FracEPlastic}
\end{figure}

First we analyze the fraction of energy remaining after traversing the coating. As shown in Fig. \ref{fig:AP8}, the flux of trapped particles is higher 
below 1 MeV so we have performed a special set of simulations for lower energies (100 and 300 KeV). As will be shown, the coating thickness is 
crucial, for that reason a lower coating thickness has also been simulated (0.15 mm). Results are shown in Fig. \ref{fig:FracEPlastic}.

As expected the energy loss increases with coating thickness. More significantly, low energy electrons and most of the protons deposit all their 
energy in the coating so that only the electrons and protons with energy greater than 1 and 10 MeV, respectively, can be detected. Electrons 
with 300 keV can be detected in case of the thinner coating.  

The energy loss for 1 MeV electrons is still quite significant (5-20\%) but decreases for higher energies. It is caused by ionization processes 
(also bremsstrahlung but less frequently) in the coating. On the other hand, protons suffer several ionizations loosing a significant fraction 
of their energy in each one. Moreover, the energy loss is still very important for 10 MeV protons that even fail to reach the scintillator in
the case of the thicker coating.

In conclusion, the thickness of the coating is a crucial parameter in the design of these experiments since it imposes a lower limit to 
the energy threshold of the assembly and severely affects the trigger rate and, moreover, in a way that depends strongly on particle type
as it has been analyzed here.

\subsection{Deposited energy in the scintillator}

\begin{figure}
\centerline{
\subfigure{\includegraphics[width=10.5cm]{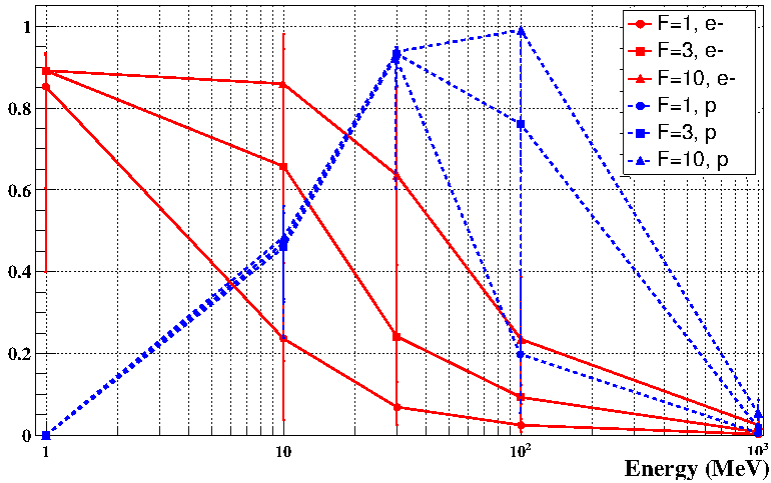}}
}
\caption{Fraction of primary energy deposited in the plastic scintillator. Error bars are the region of 68 and 95\% probability 
(in all the figures unless otherwise specified).}
\label{fig:edepplastic}
\end{figure}

Low energy electrons deposit in the plastic all their remaining energy after going through the coating, but the energy deposited decreases  
for more energetic electrons since the probability of escaping is larger as shown before in Fig. \ref{fig:edepplastic}. Again 
the different volumes of the scintillator are specially important at intermediate energies (10-100 MeV).

On the other hand, protons in the energy range from 10 to 30 MeV deposit in the scintillator all the remaining energy that they have after traversing 
the coating. 100 MeV protons deposit most of their energy in medium and large plastic volumes ($F$=3, 10) while escape from the smallest 
one ($F$=1). In case of 1 GeV, the fraction of deposited energy is only 10\% for the larger plastic and below 2\% for the smaller one.

\subsection{Energy transfer to scintillation}\label{sec:EtransfScint}

\begin{figure}
\centerline{
\subfigure{\includegraphics[width=8cm]{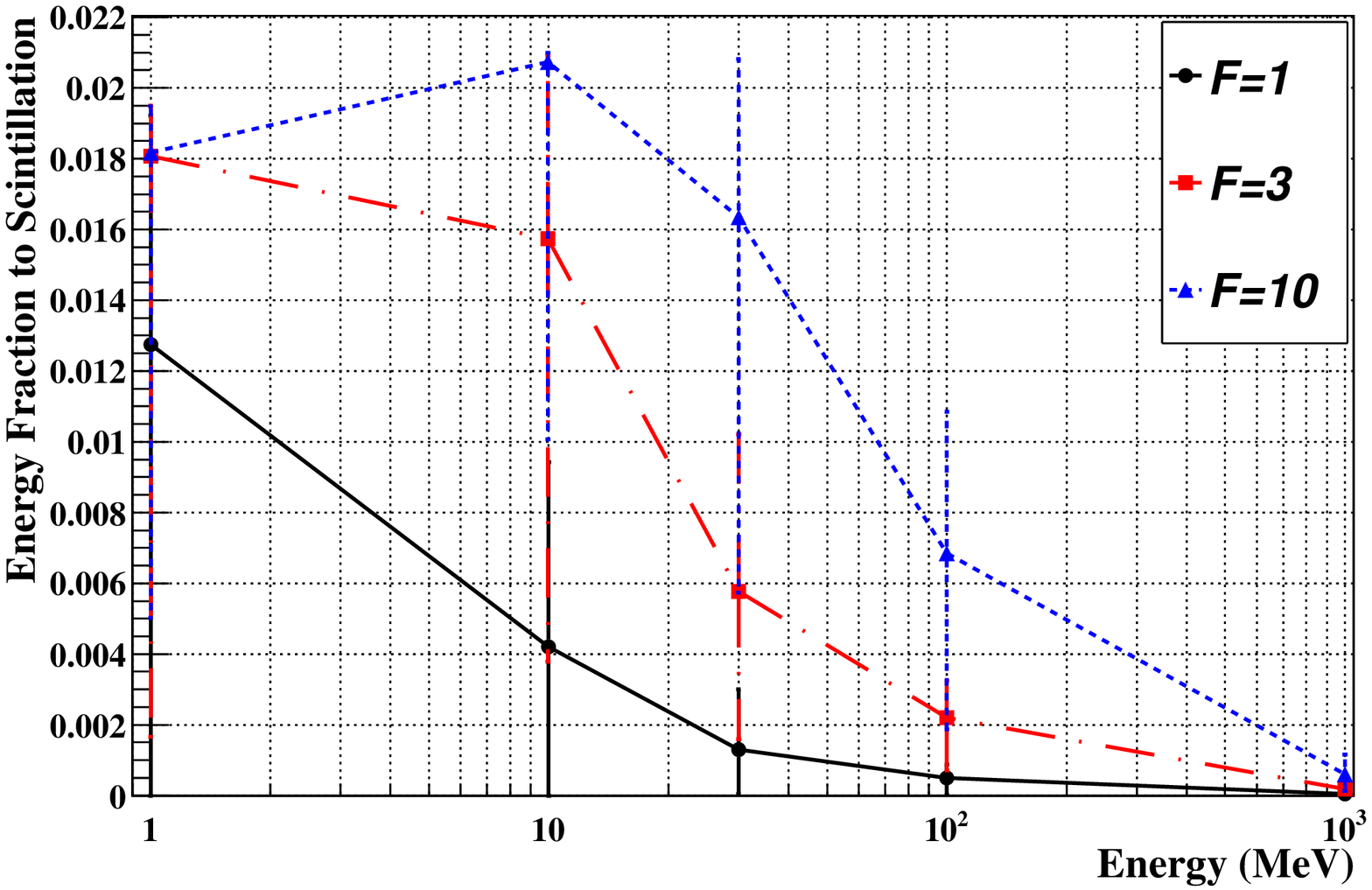}}
\subfigure{\includegraphics[width=8cm]{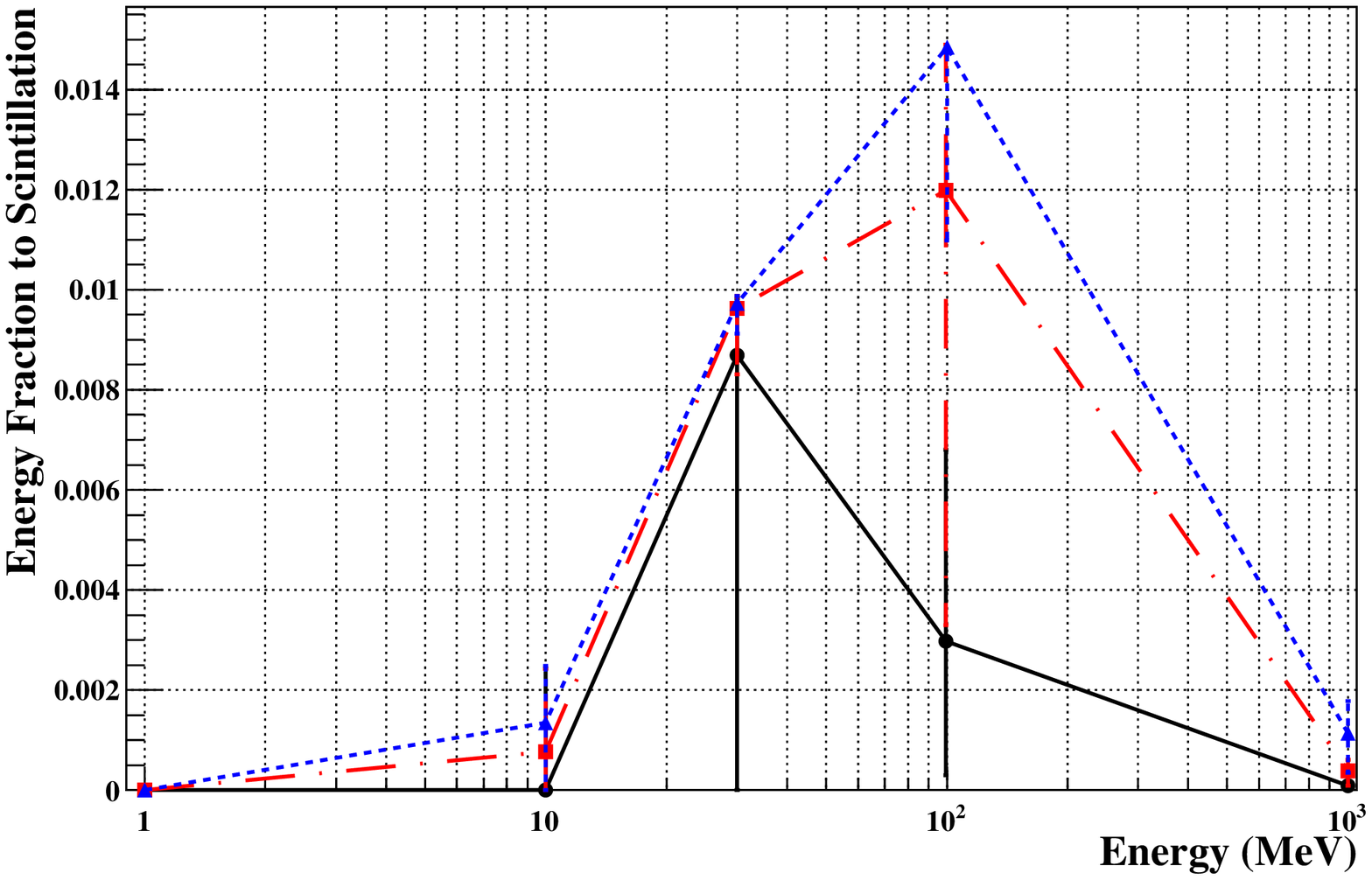}}
}
\centerline{
\subfigure{\includegraphics[width=8cm]{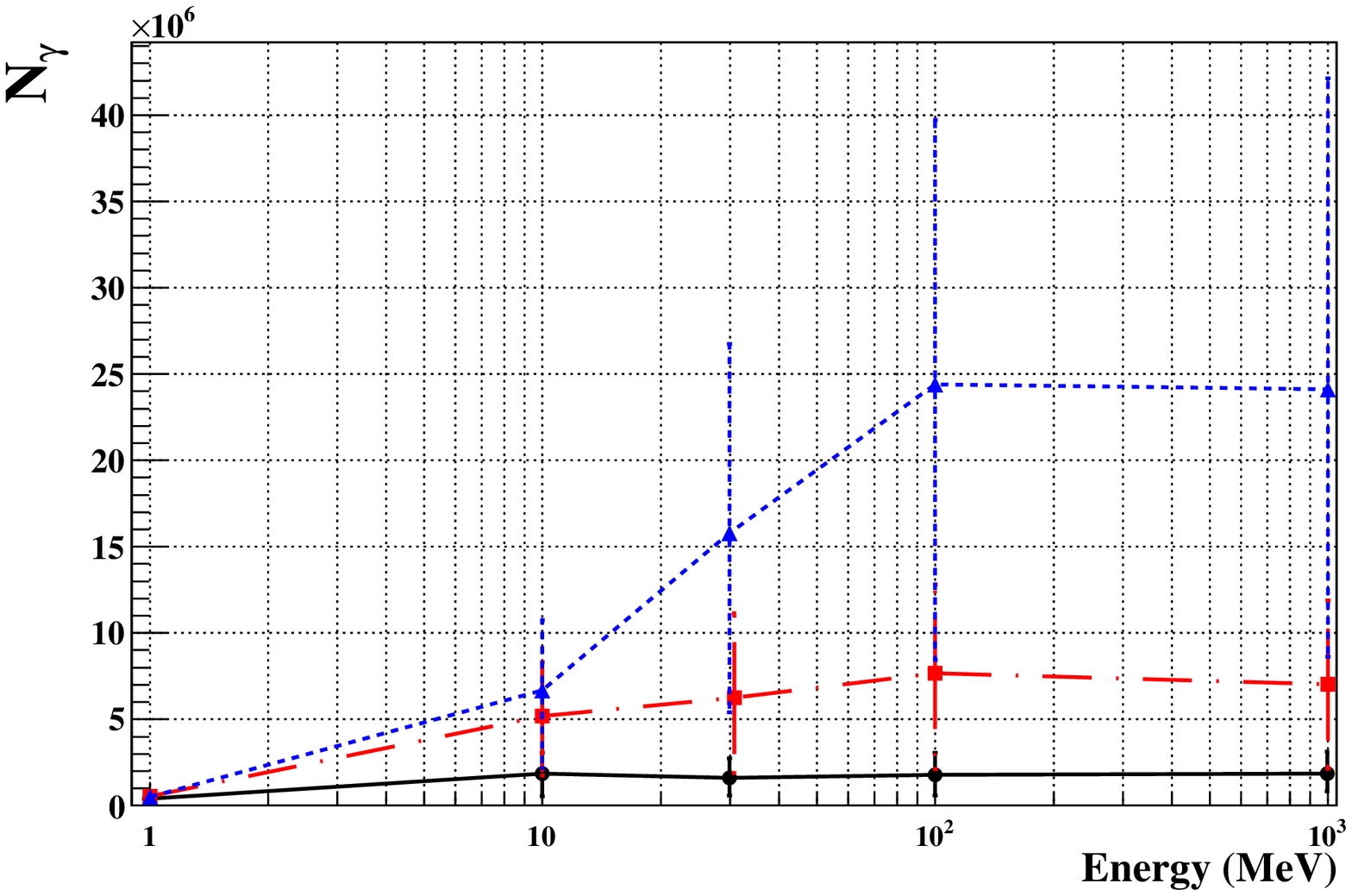}}
\subfigure{\includegraphics[width=8cm]{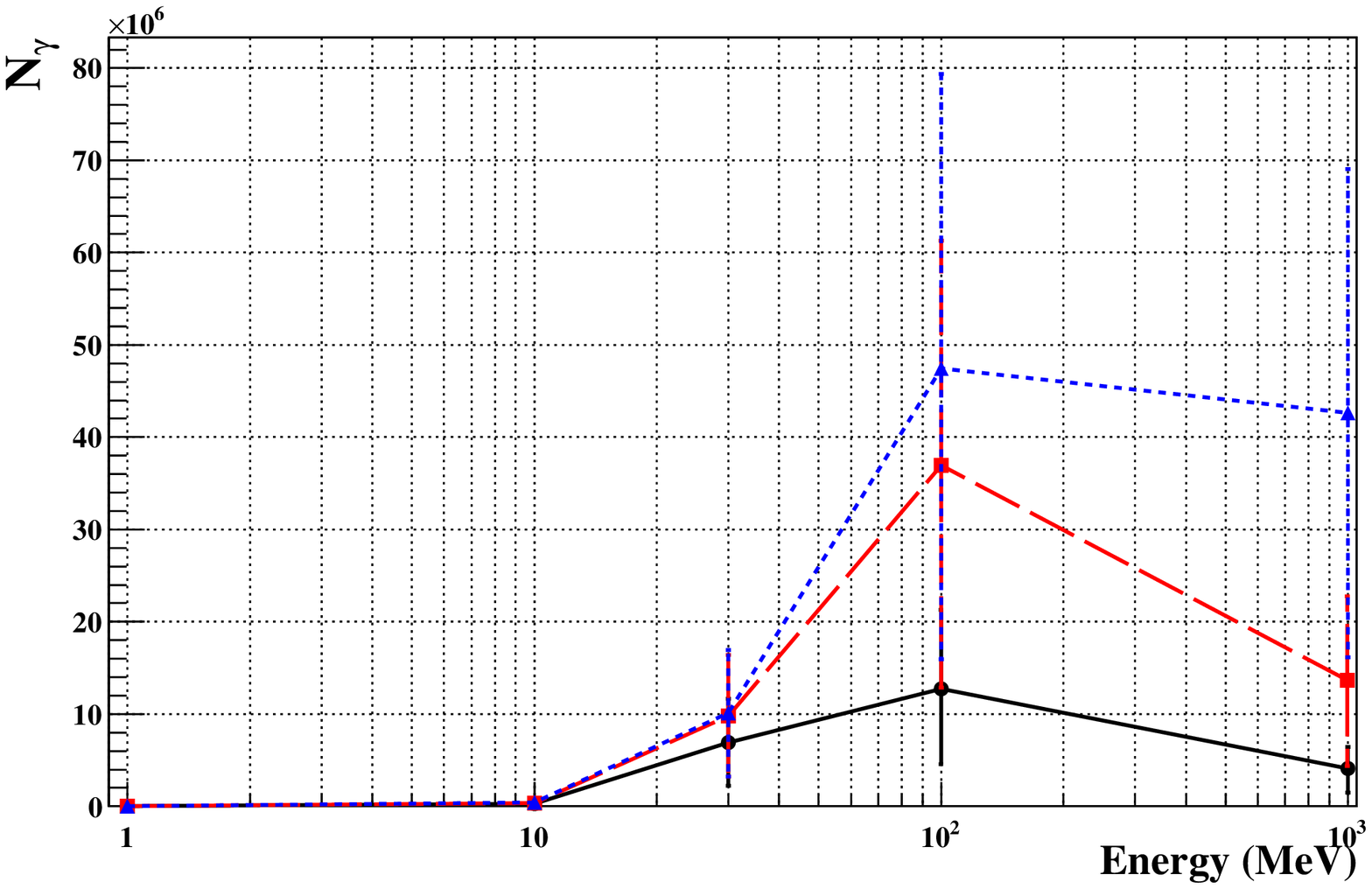}}
}
\caption{Top: Fraction of primary energy transferred to scintillation. Bottom: Number of generated photons. Left: electrons. Right: protons.}
\label{fig:scintillation}
\end{figure}

The deposited energy in the scintillator is mostly transferred to energy loss processes as ionization or bremsstrahlung and phonons and the fraction
transferred to scintillation is modeled in the simulation by the scintillation yield of the plastic. Fig. \ref{fig:scintillation}-top shows the fraction 
of the primary energy transferred to scintillation for electrons (left) and protons (right). It shows the same trend as previously shown 
in Fig. \ref{fig:edepplastic} since the scintillation yield is not dependent on the primary energy or particle type.

The result is in agreement with the known fact that the typical 
light yield (fraction of deposited energy transferred to light when the primary deposit all its energy into the plastic) in organic scintillators is 
around 2-4\%, considering that part of the energy in our case is released in the coating and the limited volume of the plastic. As can be seen, 
protons transfer less energy to scintillation as it is expected since strongly ionizing particles produce local electric fields along 
the track, which leads to quenching of scintillations, i.e., to an increasing number of non-radiative transitions in excited molecules and, 
accordingly, to a decrease in the light yield. For heavier nuclei it decreases even more strongly \cite{KharzheevReview}. Note that, as expected, 
these values are strongly dependent on the plastic volume as before. 

\begin{figure}
\centerline{
\subfigure{\includegraphics[width=8.5cm]{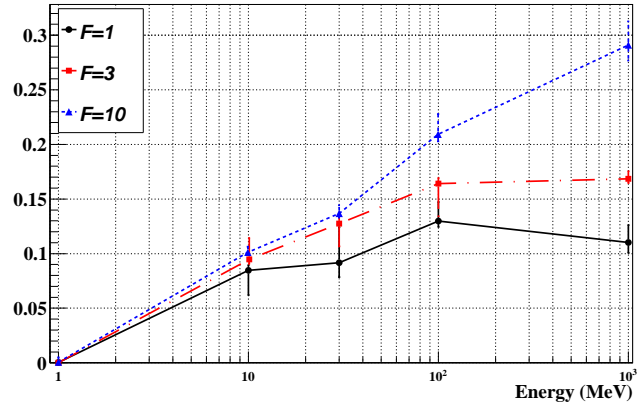}}
}
\caption{Fraction of scintillation photons produced by secondaries in case of electron primaries.}
\label{fig:scintillationsec}
\end{figure}

However, the number of scintillation photons increases with the primary energy but reaches a plateau at 100 MeV (see Fig \ref{fig:scintillation}-bottom)
or even decrease at higher energies, when the penetrating power is so high that primaries escape with less interactions. It is also interesting to 
analyze the fraction of scintillation photons produced by secondaries. In case of electron primaries this fraction increases with energy and plastic 
volume reaching values of the order of 10-25\% (Fig. \ref{fig:scintillationsec}). These photons are produced by secondary electrons of $\sim$10$^2$ KeV 
produced by ionization. On the other hand, only 1 GeV protons could produce secondary electrons by ionization (10$^2$-10$^3$ KeV) that are able 
to scintillate, producing around the 6-8\% of the scintillation photons.

The results presented here do not change significantly with thicker coating or even without it. The latter case will affect however the ability of 
the assembly to keep the light inside it and, therefore, the light pulse collected in the photo-detector as will be discussed later in 
Sections \ref{sec:Efficiency} and \ref{sec:Pulse}.

\section{Optical absorption}\label{sec:Absorption}

The scintillator absorption can produce significant light attenuation and, therefore, is the main cause of the lack of signal. First, we have 
studied the effective absorption length of our plastic, $L_{abs}$, and second, the transport efficiency due to optical absorption $\varepsilon_{att}$.
The results presented in this section do not depend on the primary energy or particle type as expected.

\begin{figure}
\centerline{
\subfigure{\includegraphics[width=8.5cm]{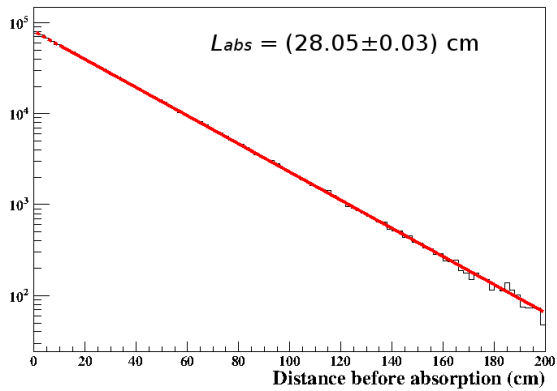}}
}
\caption{Distance traveled by optical photons before being absorbed. Exponential fit is shown. This plot corresponds to electron primaries of
10 MeV and $F=3$.}
\label{fig:AbsLenFit}
\end{figure}

Figure \ref{fig:AbsLenFit} shows the distance traveled by optical photons before being absorbed and the thick line represent and exponential fit
$Ae^{-x/L_{abs}}$. This fit is very good in all the cases analyzed (different primary types, energies and plastic volumes). We obtain
$L_{abs} \sim$  3.5, 28.0 and 81.8 cm for $F$ = 1, 3 and 10 respectively. The input value is $L_{abs}$ = 250 cm (table \ref{table:plastic}). The 
difference is due to the known fact that, in practice, there are two different attenuation lengths in scintillators. First, the bulk attenuation 
length (BAL) that depends on the scintillator material itself and how it is manufactured. This is the input value. Second, the technical attenuation 
length (TAL) that depends on the geometry and the experimental design. This is the value that is obtained from the fits. As expected, TAL increases
with plastic volume. These values are in overall agreement with measurements performed with extruded plastic in \cite{Plau-Dalmau05}. Without
coating TAL is smaller since the probability of escaping the plastic is much larger. We obtain $L_{abs} \sim$  1.2, 7.5 and 23.0 cm for $F$ = 1, 3 and 10 respectively in this case. 

It is interesting that it is possible to obtain from the simulation the bulk attenuation length used as input. To that end, a special simulation setup 
must be performed to avoid any reflections in the borders (for example throwing the primaries straight to the scintillators) and assuring that 
all of the photons will be absorbed (with a very large volume of plastic), as for example it is done in \cite{DietzLaursonThesis, Riggi}.

Regarding $\varepsilon_{att}$, it is determined from the fraction of scintillation photons that are not absorbed in the plastic. It corresponds to
98.5, 88.7 and 77.1\% for $F=$ 1, 3 and 10 respectively. It decreases when increasing the volume as expected. Without coating, $\varepsilon_{att}$ 
is higher (92.3, 97.0 and 90.5\%) because many photons escape from the plastic in this case as will be analyzed later (Section \ref{sec:Efficiency}).

\section{Light Collection}\label{sec:Efficiency}

In this section we will analyze the light collection of the assembly. Photons could reach the photo-detector directly from its generation point
or after being reflected one or several times in the coating. The former represents the geometrical collection factor of the detector 
($F_{geom}$) and the latter the collection factor due to reflections ($F_{ref}$), both calculated as the fraction with respect to the total 
number of photons produced. 

\begin{figure}[h]
\centerline{
\subfigure{\includegraphics[width=7.5cm]{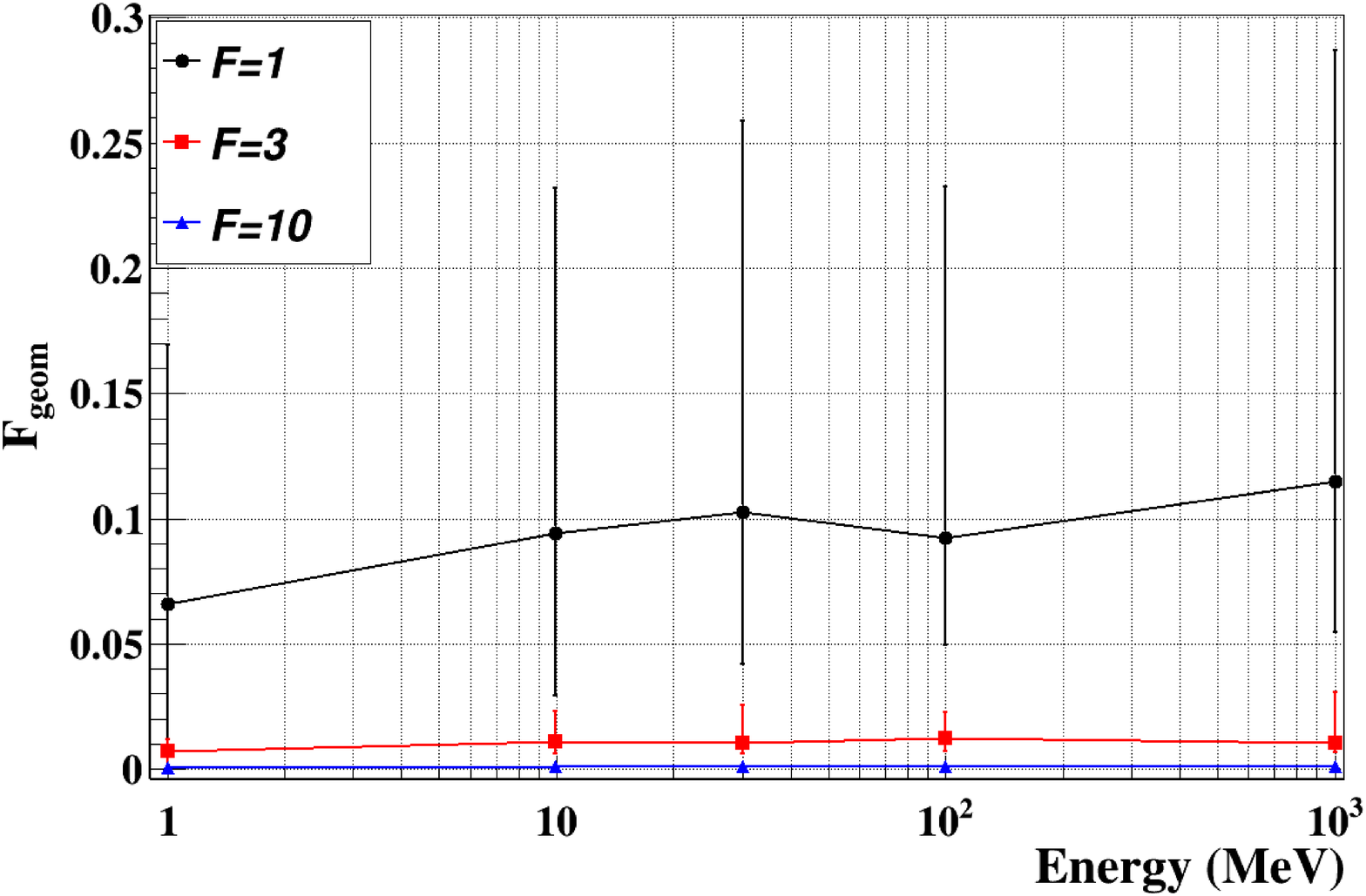}}
\subfigure{\includegraphics[width=7.5cm]{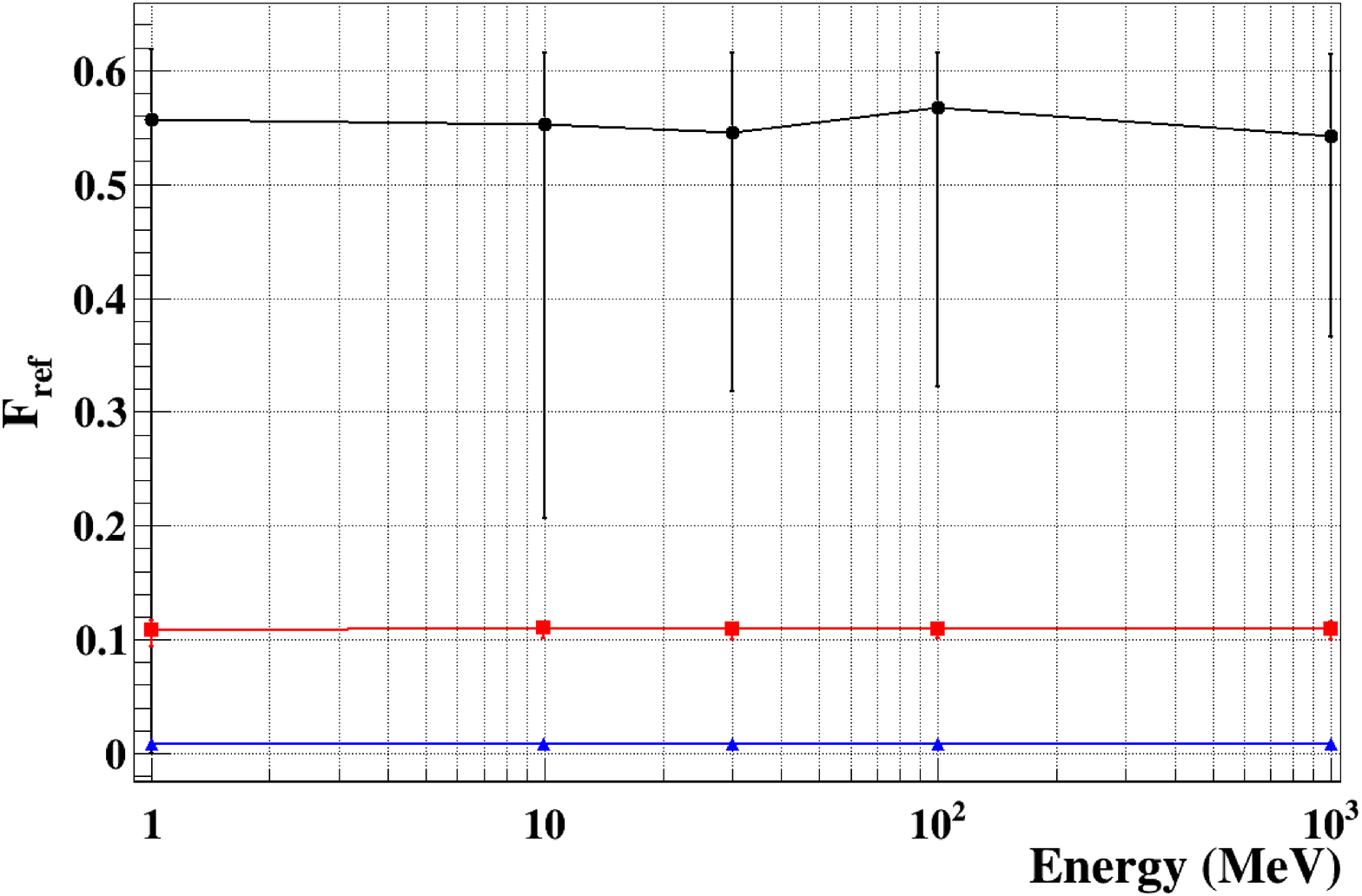}}
}
\caption{Geometrical collection factor ($F_{geom}$) and collection factor due to reflections ($F_{ref}$) as a function of the primary energy for
different plastic volumes.}
\label{fig:factors}
\end{figure}

The results are shown in Fig. \ref{fig:factors}. $F_{geom}$ is negligible for the largest plastic volume and only around 0.8\% for the intermediate one, 
while it represents $\sim$8\% of the total number of photons produced in case of the smallest scintillator. Regarding to $F_{ref}$, it is 
$\sim$56, 10 and 0.1\% for $F$= 1, 3 and 10 respectively, so most of the light is detected after one or several reflections in the plastic-coating 
interface. The result is obviously independent of primary type or energy.  

Without coating, $F_{geom}$ obviously do not change, but $F_{ref}$ is reduced to $\sim$25, 2.3 and 0.2\% for $F$= 1, 3 and 10 respectively, since 
the probability of reflection in the plastic-vacuum interface is much lower. Therefore, the collected light is greatly enhanced thanks to the 
reflective coating. 

\begin{figure}[h]
\centerline{
\subfigure{\includegraphics[width=8.5cm]{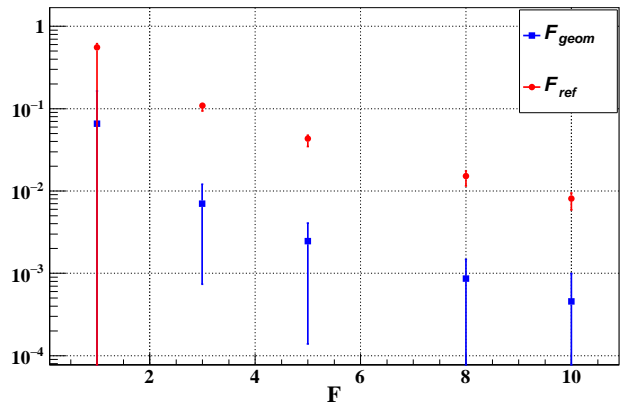}}
}
\caption{Geometrical collection factor ($F_{geom}$) and collection factor due to reflections ($F_{ref}$) as a function of the plastic size ($F$).}
\label{fig:factors2}
\end{figure}

To further investigate the dependence of  $F_{geom}$ and $F_{ref}$ with plastic volume, an independent set of simulations for 1 MeV electrons have 
been performed for $F$=5 and 8. As can be seen from Fig. \ref{fig:factors2} both decrease with $F$. The reason for the behavior of $F_{geom}$ is that 
the solid angle subtended from the collection area of the photo-detector decreases strongly when increasing the plastic volume. The trend of 
$F_{ref}$ is due to the photons lost in the reflections in the plastic-coating interface as it is analyzed in the next section. 

\section{Lost photons}\label{sec:lostphotons}

In order to quantify the loss of photons and the strong decrease of $F_{ref}$ with plastic volume, we define the coefficients R, RT, A and T 
that represent the fraction of the times that a photon in the plastic-coating (or plastic-vacuum) interface suffers reflection, total 
internal reflection, it is absorbed or transmitted, respectively. 

\begin{figure}[h]
\centerline{
\subfigure{\includegraphics[width=8.5cm]{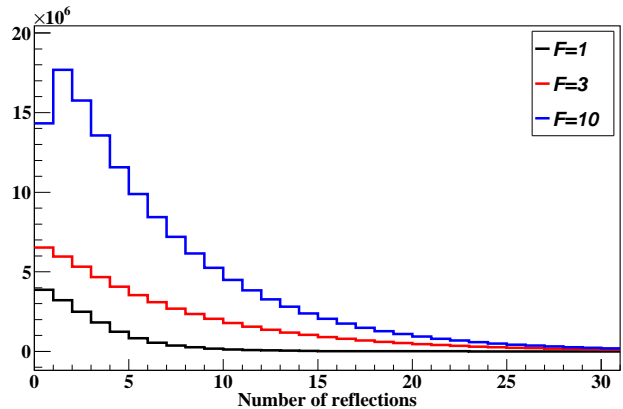}}
}
\caption{Number of reflections for the different plastic volumes considered.}
\label{fig:reflections}
\end{figure}

In the case with coating, these values are 90, 0, 10 and 0\% respectively, reproducing correctly the properties imposed to this interface 
in Section \ref{sec:Surfaces} and according to the index of refraction of the materials (n$_{coating}$ $>$ n$_{plastic}$, so internal reflection 
is not possible). The 10\% probability of absorption is the responsible for the strong decrease of $F_{ref}$ with plastic volume since the number
of reflections increases strongly for larger plastic volumes. It is shown in Fig. \ref{fig:reflections} where the distribution of the number 
of reflections for each photon produced in all the simulations is shown. As commented in Section \ref{sec:Surfaces} this absorption is actually 
a consequence of the fact that photons will be transmitted to the coating whose attenuation length is so small that will be optically absorbed 
inside it.

Without coating, we obtain R=4.3\%, RT=60\%, A=0\% and T=35.7\%. Therefore, the probability of a photon to escape each time it reaches the edge
of the plastic is 35.7\% without coating compared with only 10\% of absorption probability in the plastic-coating interface with it. 
Therefore, much more light is retained thanks to the coating. These probabilities do not change for different plastic volumes since only depend 
on the material properties. Thus, the total fraction of photons lost compared to the number of photons produced without coating is 62, 94.5 and 99.7\% 
for $F$=1, 3 and 10 respectively. Again, the benefit of using the coating is evident as it will also be proved in Sec. \ref{sec:Pulse} when 
the light pulse at the photo-detector are studied. 

It is also interesting to note that Cerenkov photons are produced by electrons inside the coating and the plastic. They represent a very low fraction 
compared to scintillation photons. Those produced in the coating suffer several reflections inside the coating but most of them finally enter 
to the plastic since T is only 15\% in the coating-vacuum interface (the rest suffer total internal reflection).

The results presented in this section do not depend on the thickness of the coating since reflectivity has been considered the same for the two
cases analyzed (0.25 and 0.50 mm). In practice, it is likely that reflectivity increases slightly with thickness since TiO$_2$ fraction is
constant in weight. This will be analyzed later in Sec. \ref{sec:reflectivity}.

\section{Pulses at the photo-detector}\label{sec:Pulse}

The fraction of photons that are detected with respect to the total of them produced by scintillation is 70, 12 and 0.8\% with coating 
for $F$= 1, 3 and 10 respectively. The significance of the plastic volume is evident. Without coating, the fraction is 38, 3.5 and 0.3\% 
so the use of coating increases the collected light by a factor 2-3. Next the shape of the pulse and its time parameters will be analyzed.

\subsection{Pulse shape}
\begin{figure}[h]
\centerline{
\subfigure{\includegraphics[width=6.0cm]{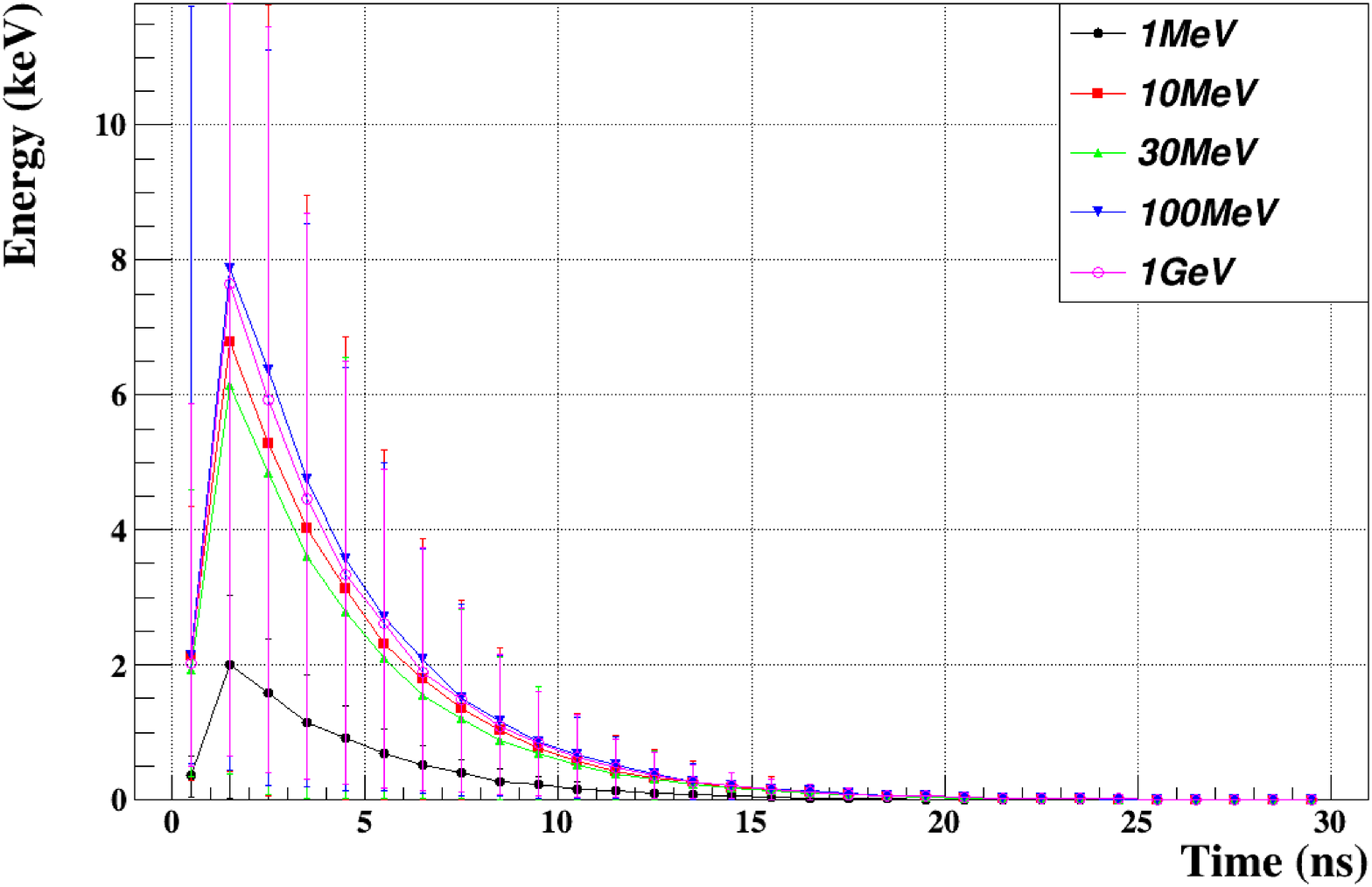}}
\subfigure{\includegraphics[width=6.0cm]{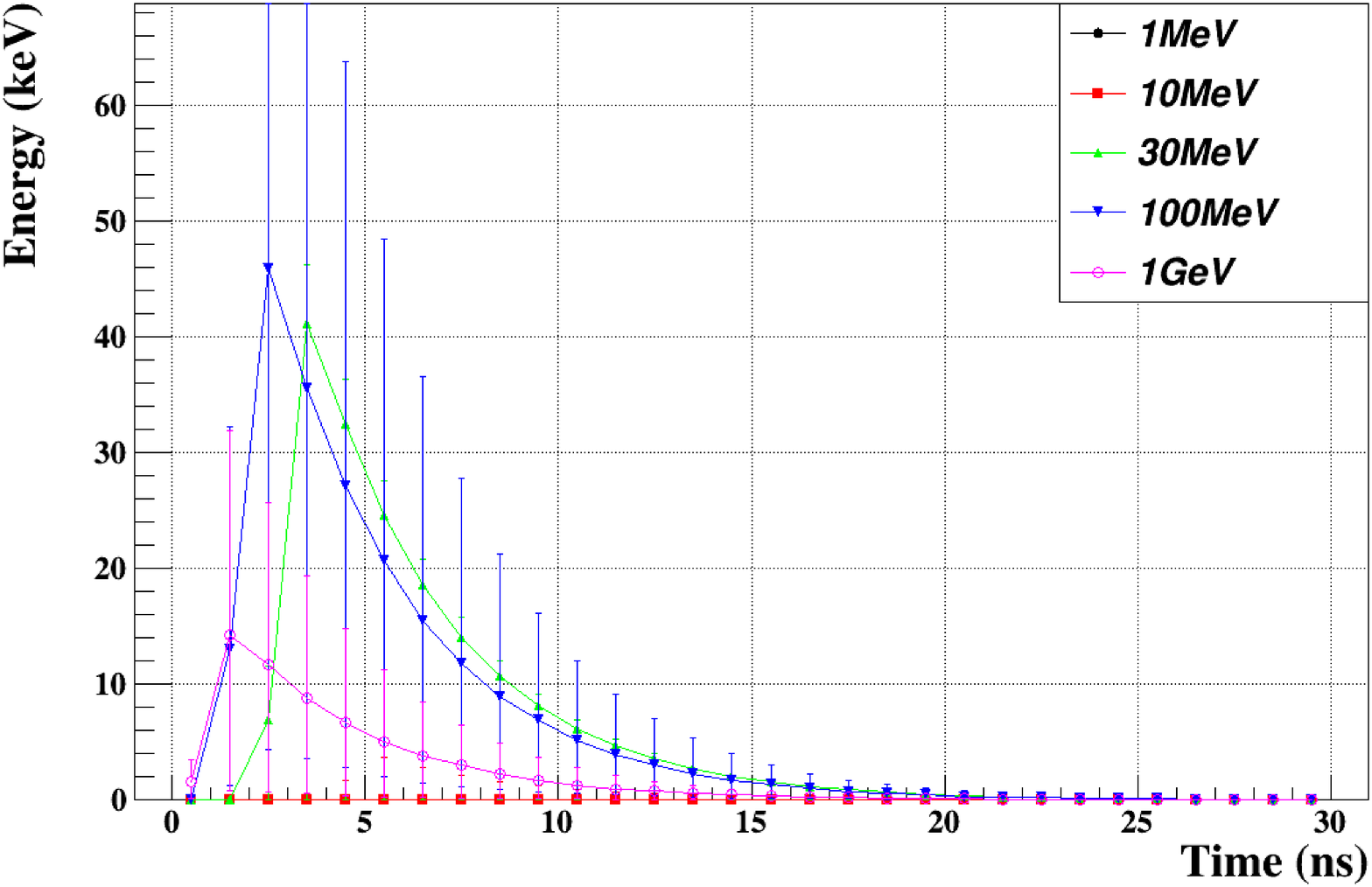}}
}
\centerline{
\subfigure{\includegraphics[width=6.0cm]{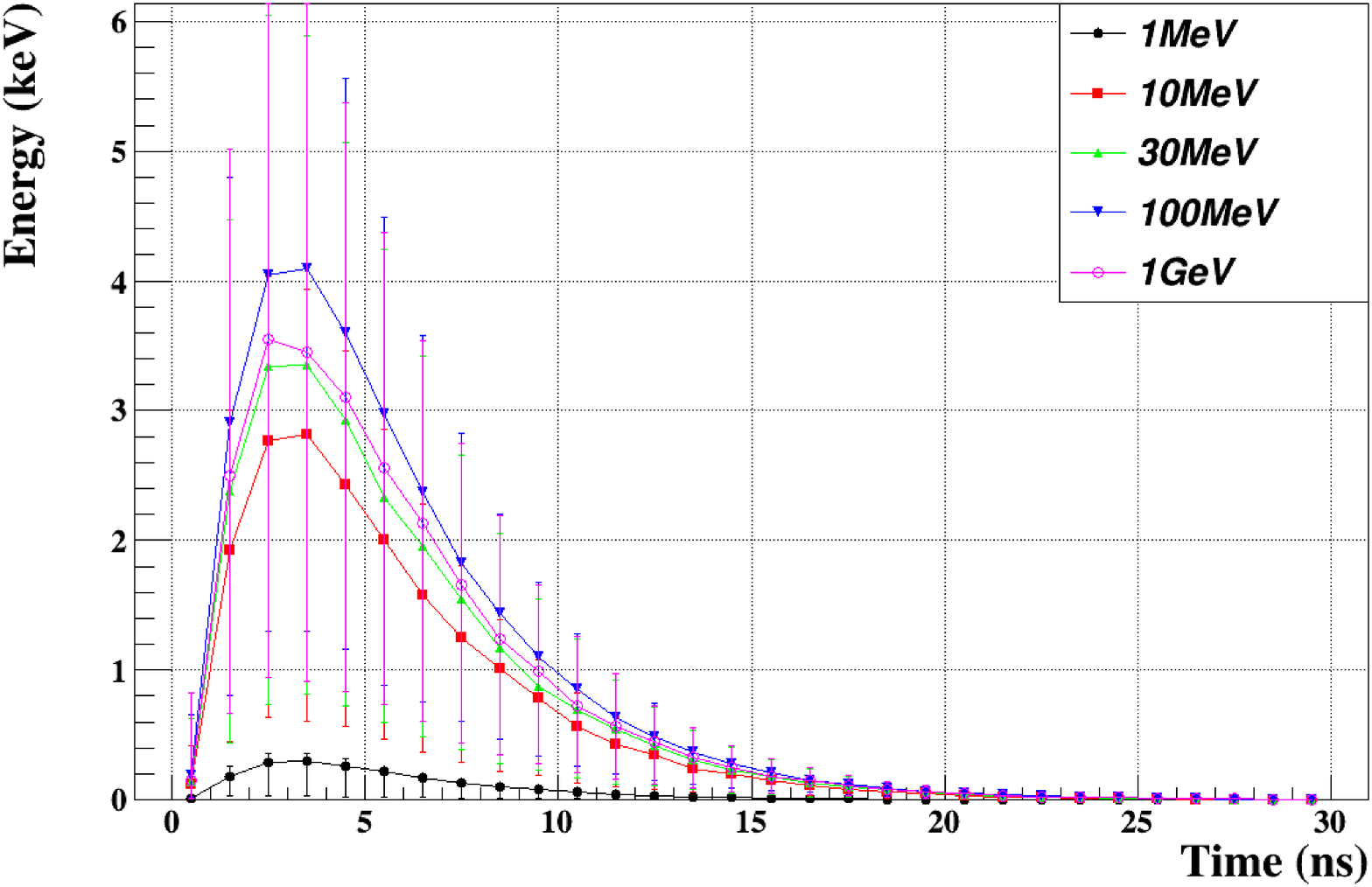}}
\subfigure{\includegraphics[width=6.0cm]{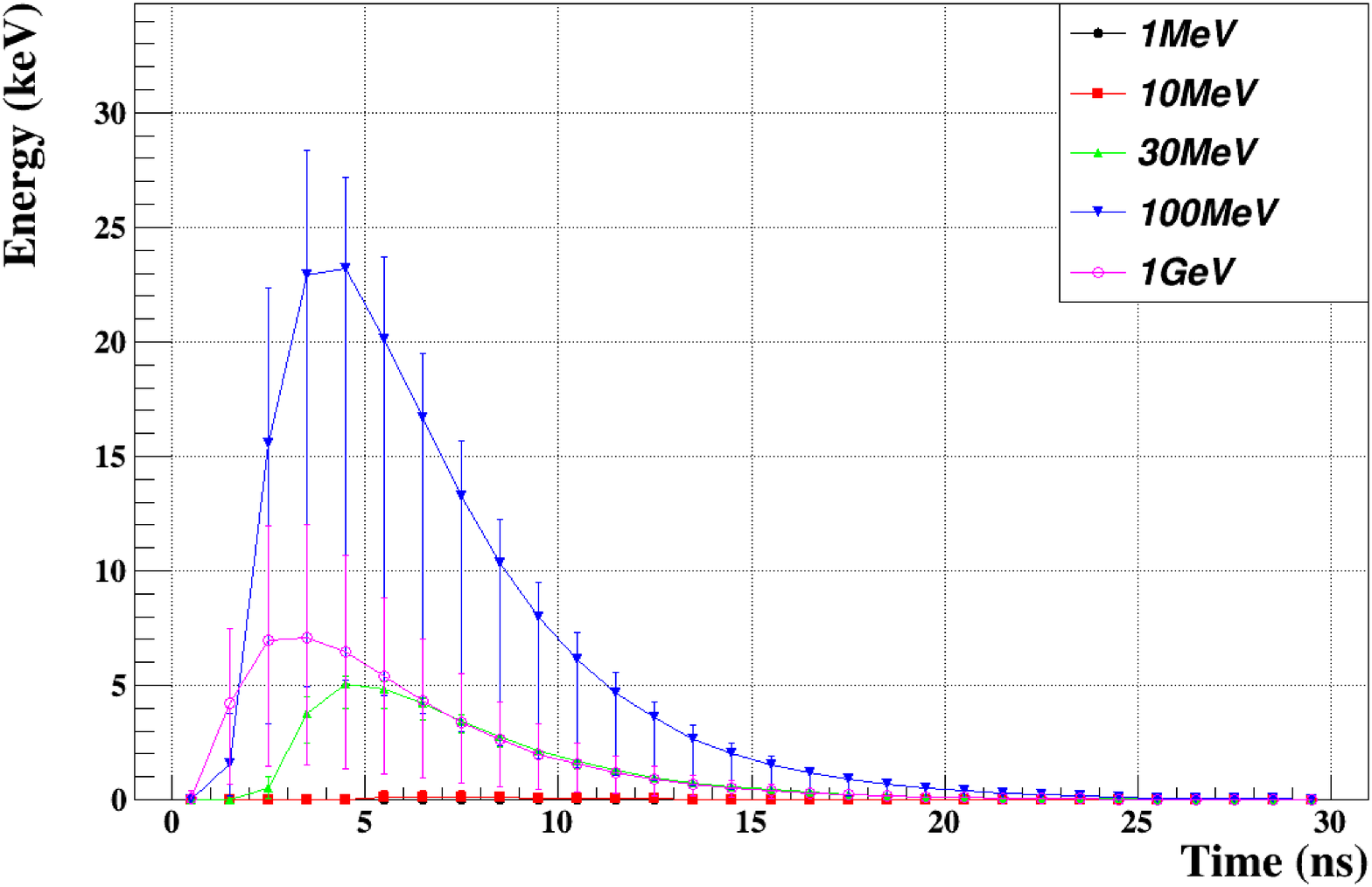}}
}
\centerline{
\subfigure{\includegraphics[width=6.0cm]{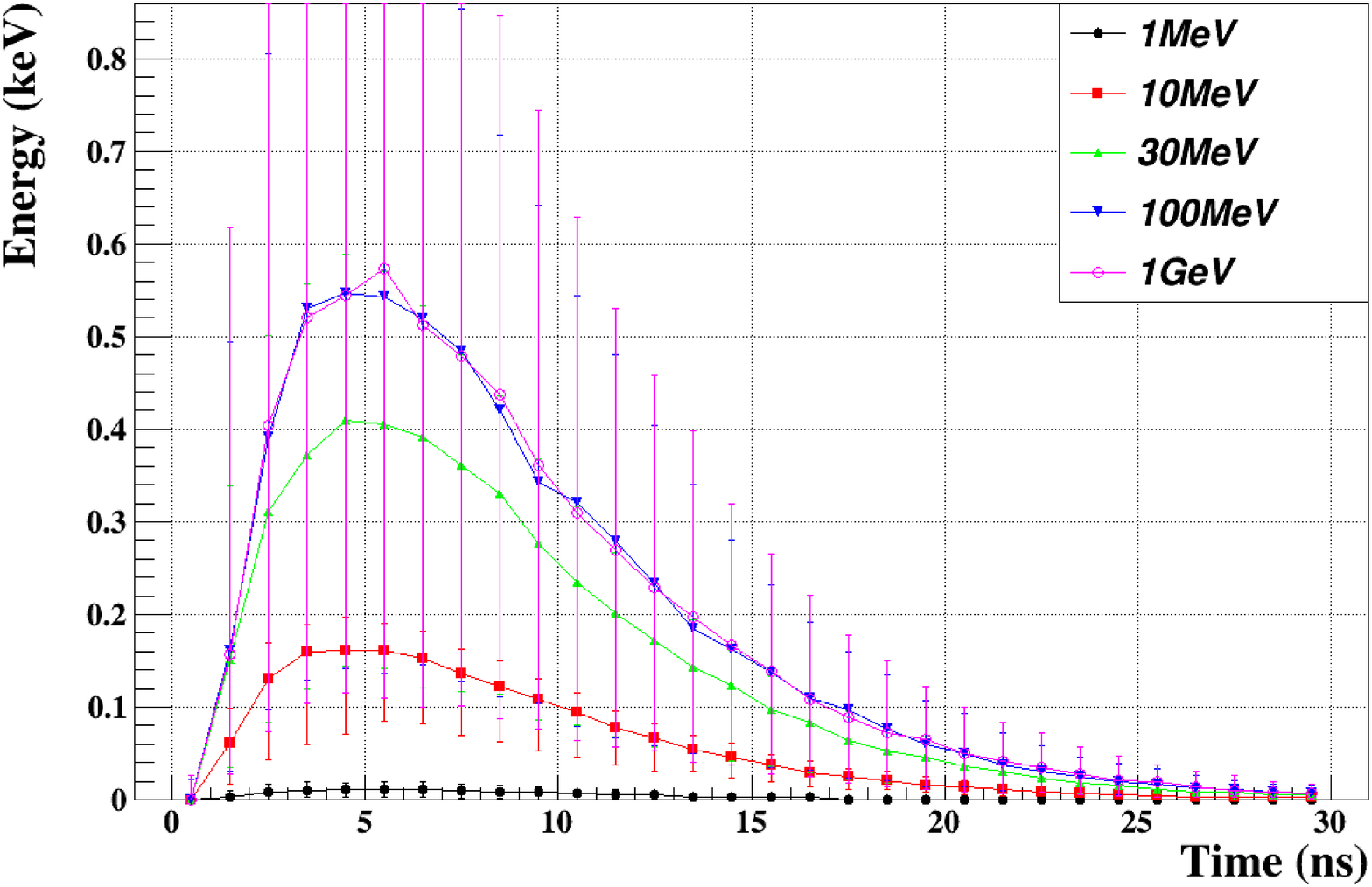}}
\subfigure{\includegraphics[width=6.0cm]{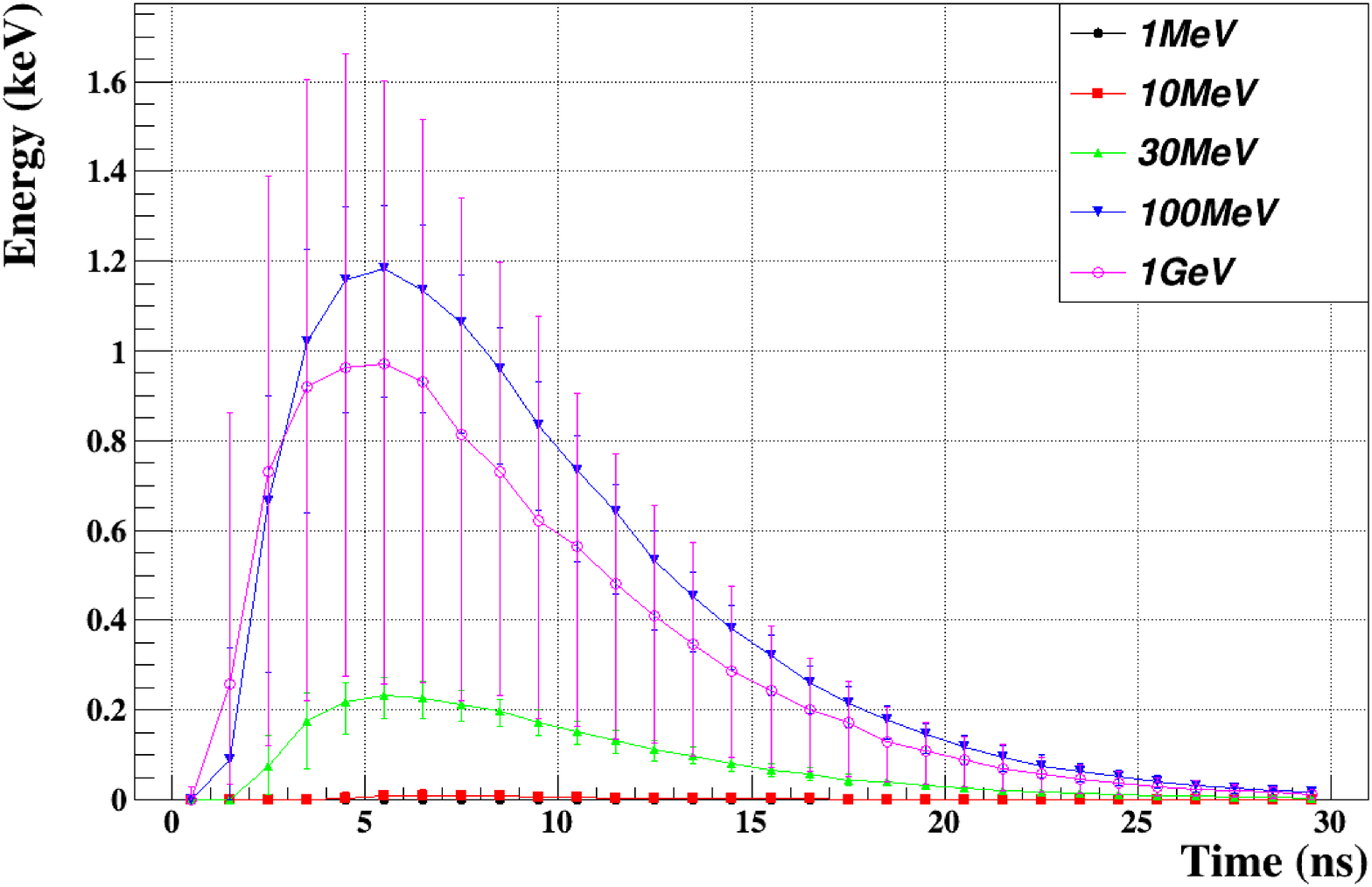}}
}
\centerline{
\subfigure{\includegraphics[width=6.0cm]{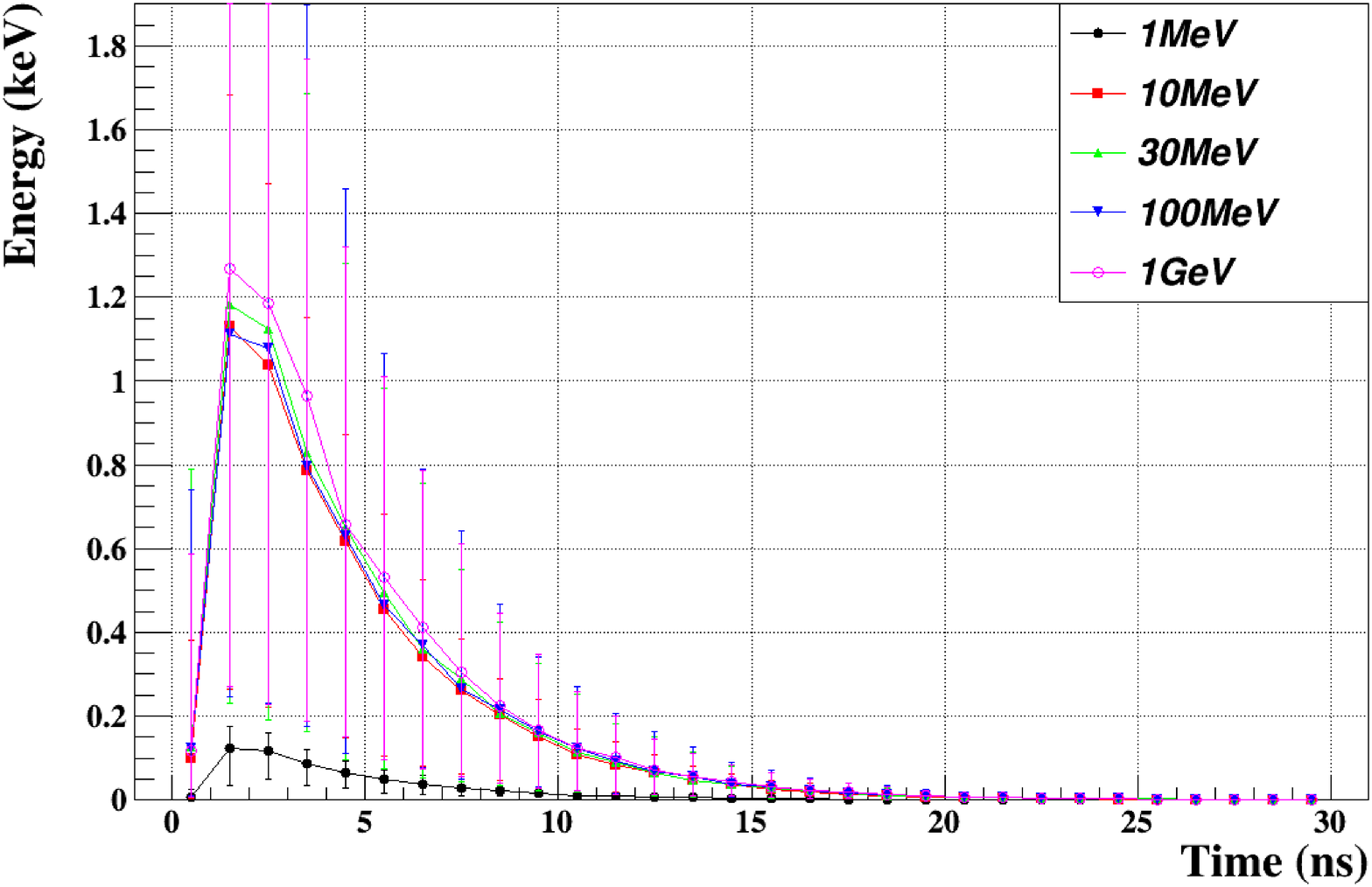}}
\subfigure{\includegraphics[width=6.0cm]{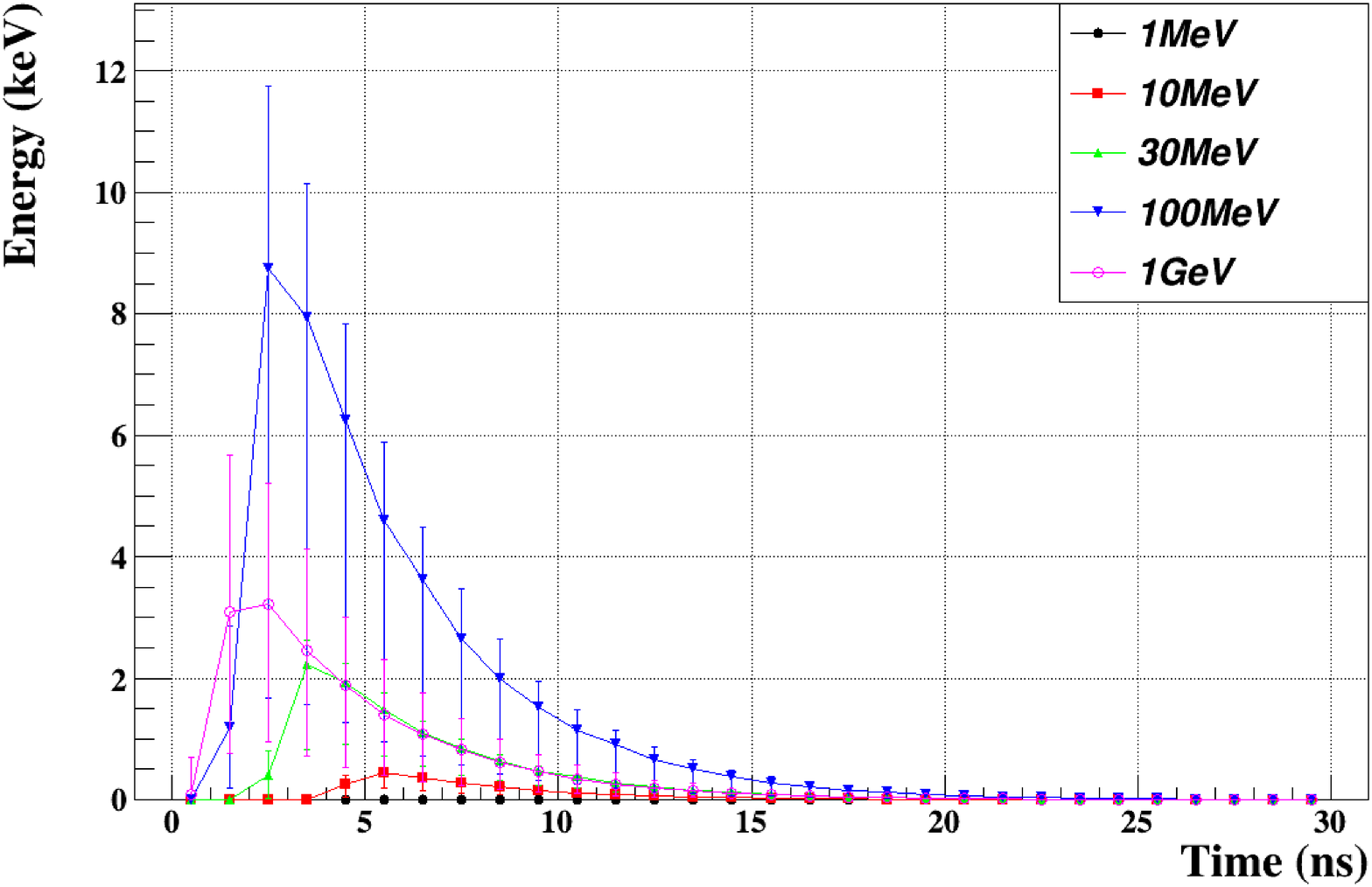}}
}
\caption{Pulses in the photo-detector. Left: electrons. Right: protons. From top to down: $F$=1, 3 and 10. The last row corresponds to $F$=3 
without coating. The points represent the median and the error bars the 68\% region of probability of the 200 simulations performed.}
\label{fig:pulses}
\end{figure}

\begin{figure}[h]
\centerline{
\subfigure{\includegraphics[width=7.5cm]{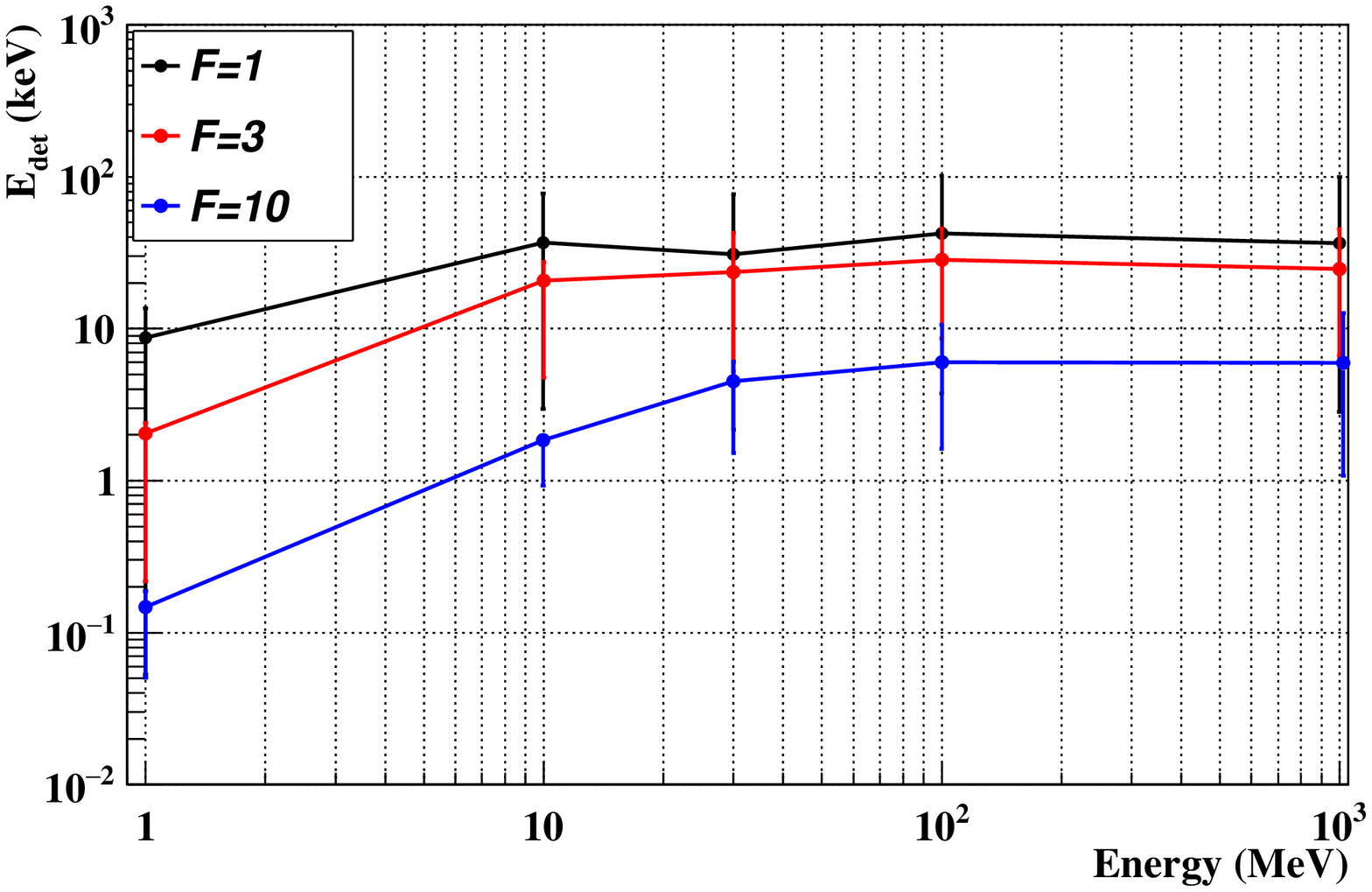}}
\subfigure{\includegraphics[width=7.5cm]{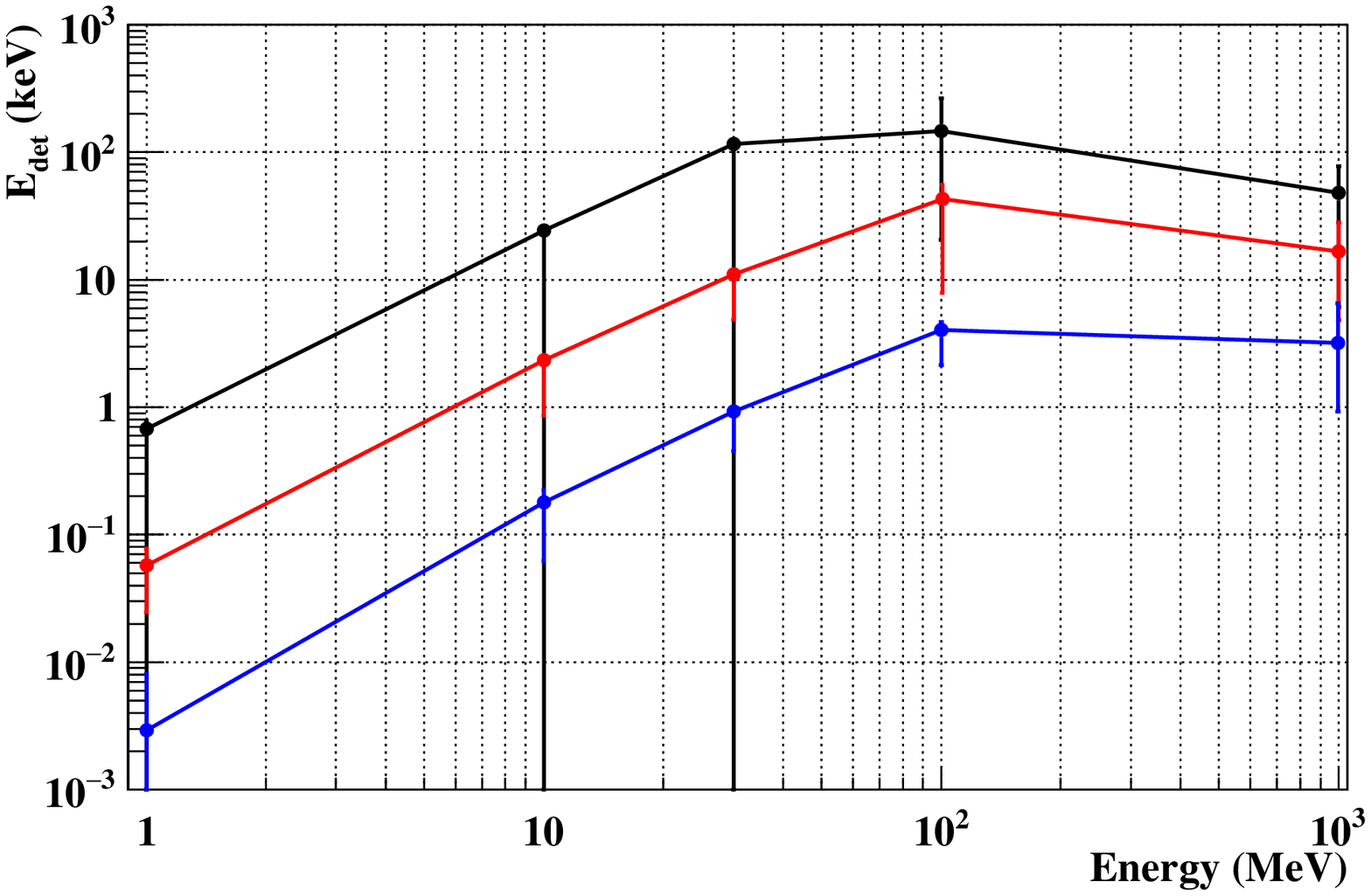}}
}
\centerline{
\subfigure{\includegraphics[width=7.5cm]{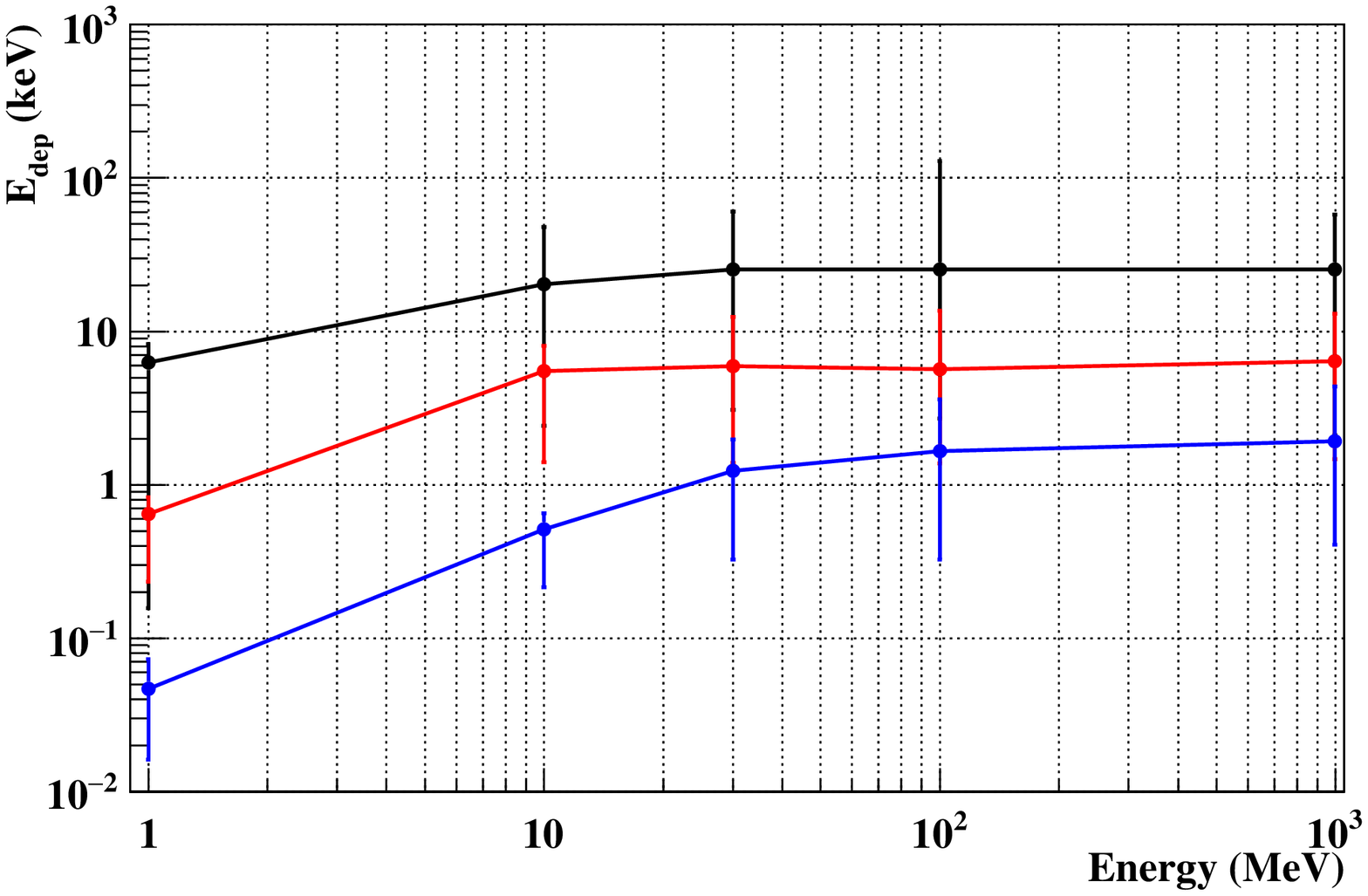}}
\subfigure{\includegraphics[width=7.5cm]{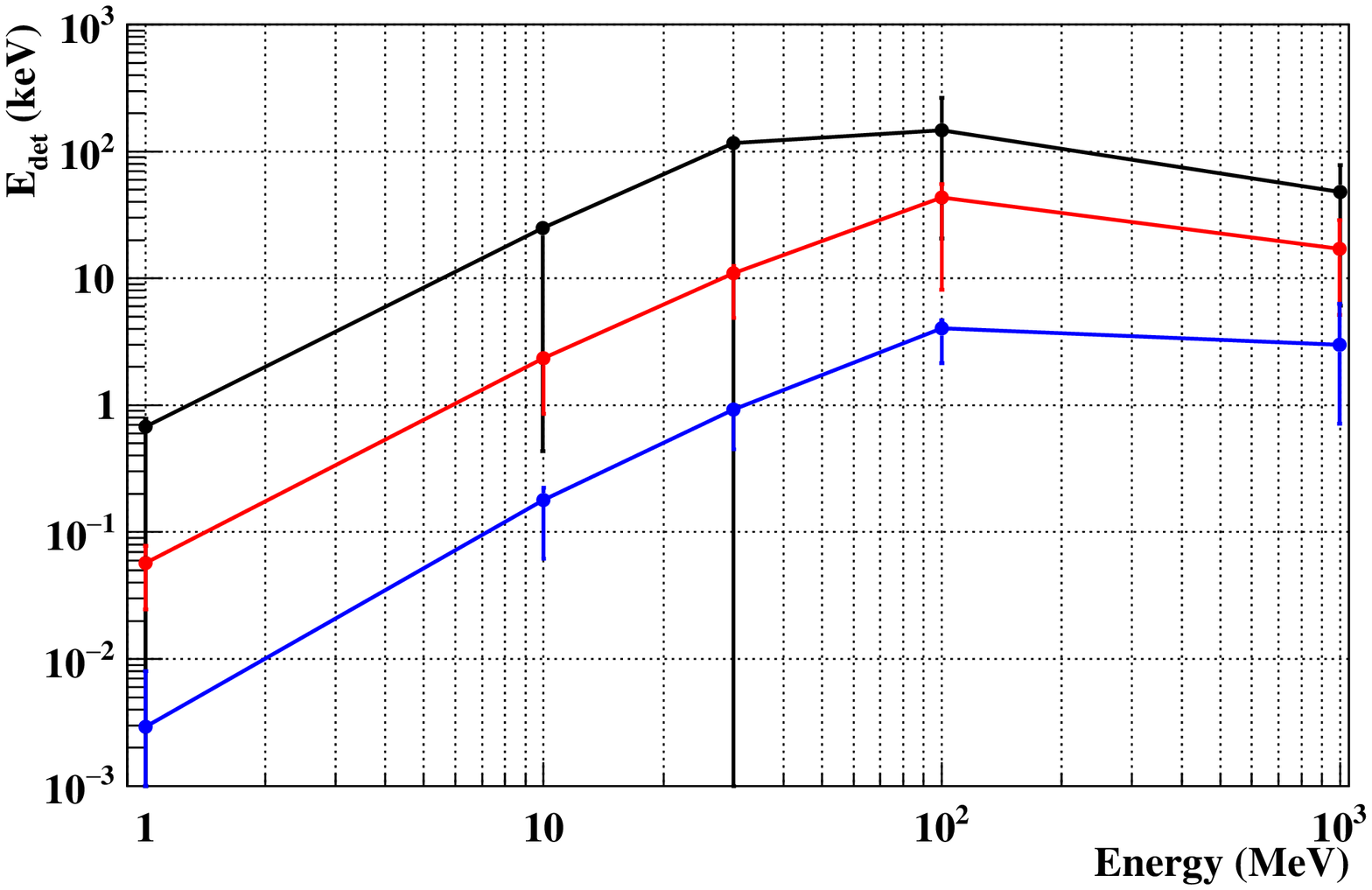}}
}
\caption{Energy of the integrated pulse as a function of the primary energy for $F$=1, 3 and 10, with (top) and without (bottom) coating. 
Left: electrons. Right: protons.}
\label{fig:totalsignal}
\end{figure}

In Fig. \ref{fig:pulses} the light pulse at the photo-detector is shown for electron and proton primaries and the primary energies 
and plastic volumes considered. The case without coating is shown for $F$=3 in the last row. Fig. \ref{fig:totalsignal} shows the integrated 
signal at the photo-detector, $E_{det}$. Several remarks can be done:
\begin{itemize}
 
 \item The pulse starts earlier for electron than for proton primaries of the same energy since electrons travel faster inside 
 plastic than protons do.
 \\ 
 \item The time of the pulse maximum is almost independent of the electron energy since they scintillate almost immediately after entering 
 into the plastic and scintillate in the whole volume. On the contrary, this time decreases when increasing the proton energy since they 
 are more penetrating as energy increases and then scintillates closer to the photo-detector. 
 \\
 \item The pulse is wider and the peak less pronounced as plastic volume increases since photons arrived later at the photo-detector. Several 
 time parameters will be analyzed later.
 \\
 \item The larger the plastic volume, the easier it is to estimate the primary energy using the signal of the peak of the pulse. 
 However, the regions of 68\% probability shown are very large (more for higher energies and small volumes) so the energy or particle type 
 discrimination on event by event basis is very unlikely. This is caused, mainly, because primary flux is isotropically arriving to the assembly 
 with particles crossing it to a great extent while others doing it only through a corner. Obviously, this fact is more important for smaller plastics but
 also for higher energies since they could easily cross the entire plastic producing photons along their track. 
 \\
 \item Although a large number of photons are produced as plastic volume increases (see Fig \ref{fig:scintillation}-bottom), the integrated 
 signal decreases (Fig. \ref{fig:totalsignal}). A similar behavior is found and discussed in \cite{Archambault}. The integrated signal is reduced  
 40-60\% and 80-95\% as plastic volume increases from $F$=1 to 3 and 10 respectively. As discussed before, the main reason 
 is that as volume increases, also does the number of reflections in the coating surface where 10\% of absorption probability exists (see Sec.
 \ref{sec:lostphotons}). This is enhanced by the fact that the optical absorption also increases for larger plastics (Sec. \ref{sec:Absorption}). 
 \\
 \item The total signal is significantly larger for proton primaries of 30-100 MeV than electrons of the same energy. This is expected as 
 explained in Sec. \ref{sec:EtransfScint}. Protons at low energies barely enter the plastic while the most energetic ones escape from the plastic
 so they are not directly comparable with electrons.
 \\
 \item Not only scintillation photons are detected but Cerenkov photons as well. They are produced in the co-extruded coating and in the plastic. 
 They arrive at the detector in the first 0.5-3 ns (depending on the plastic volume). However they represent a negligible part of the signal, 
 always $\leq$ 2\% and, in case of proton primaries with E $\leq$ 1GeV they are not even produced. In contrast, its undesirable effect 
 is very important in dossimetry \cite{Archambault}.
 \\
 \item Without coating the pulse is narrower and the peak occurs earlier since many photons escape from the plastic instead of arriving 
 at the photo-detector after several reflections. The integrated signal is reduced between 30 to 80\% depending on particle type, energy 
 and plastic volume. 
 
\end{itemize}

\subsection{Time parameters of the pulse}\label{sec:TimeParam}

\begin{figure}[h]
\centerline{
\subfigure{\includegraphics[width=7.5cm]{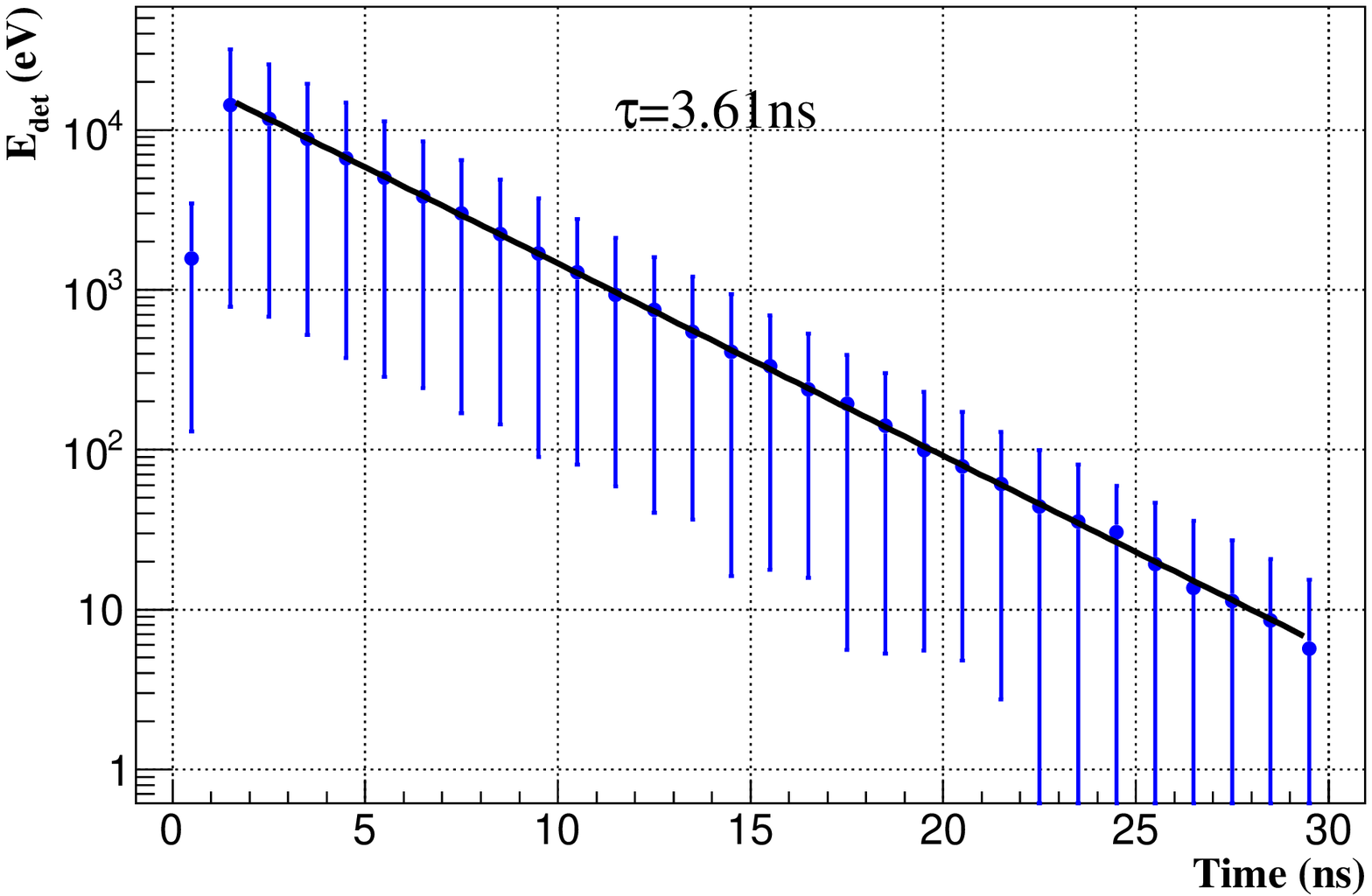}}
\subfigure{\includegraphics[width=7.5cm]{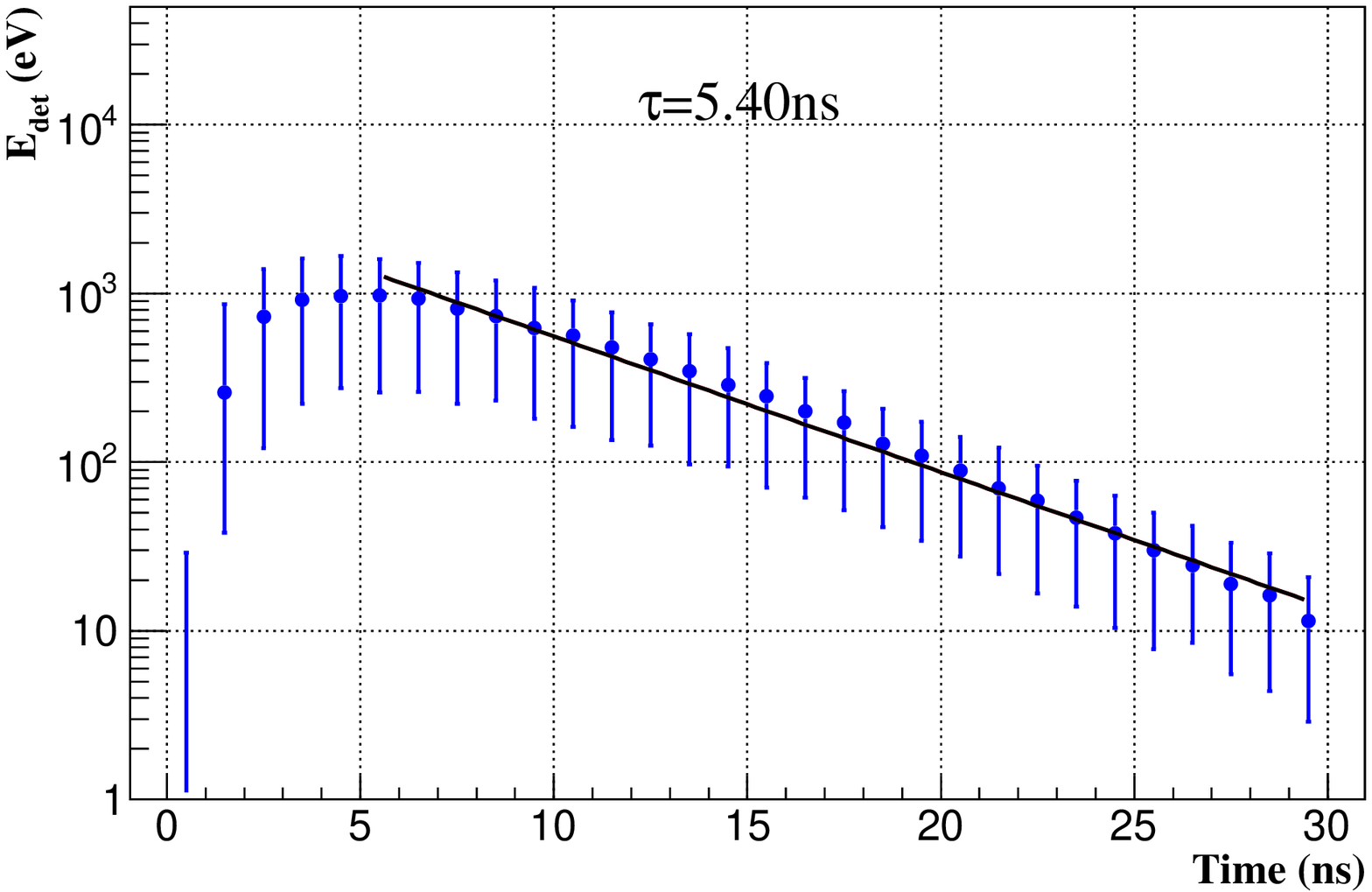}}
}
\caption{Fit to the tail of the pulses for 1 GeV protons and the plastic volume $F$=1 (left) and 10 (right).}
\label{fig:timedecay}
\end{figure}

The time parameters associated to the pulse shape are analyzed. First, an exponential fit, $Ne^{-t/\tau}$, is performed to the tail 
of the pulses (starting at the time of the maximum). The value set in the simulation for the decay response of the 
scintillator is perfectly reproduced for every energy and primary for smaller plastics as shown in Fig. \ref{fig:timedecay}-left. However,
the pulse is wider as the plastic volume increases since photons arrive at the photo-detector after many reflections, so an exponential 
do not describe so well the pulse decay (see Fig. \ref{fig:timedecay}-right). 

\begin{figure}[h]
\centerline{
\subfigure{\includegraphics[width=8.5cm]{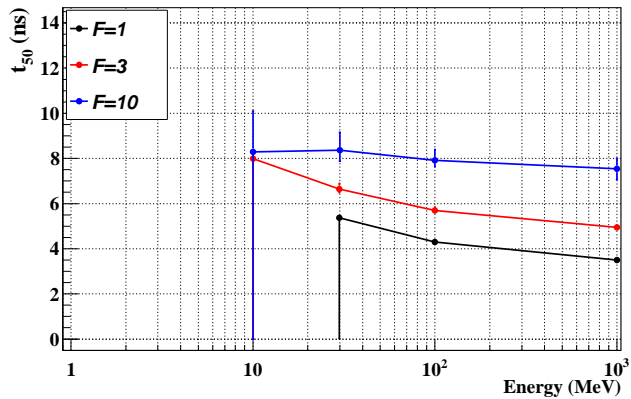}}
}
\caption{t$_{50}$ for proton primaries as a function of energy for $F$=1, 3 and 10.}
\label{fig:t50proton}
\end{figure}

Next, the time when the signal reaches 10 (t$_{10}$), 50 (t$_{50}$) and 90\% (t$_{90}$) of its integrated value are calculated. This
values are energy independent in case of electron primaries. Results are shown in Table \ref{table:summary}. In case of protons, 
the time parameters vary slightly with energy as it is shown for t$_{50}$ in Fig. \ref{fig:t50proton}. This can be understood from
Fig. \ref{fig:pulses}, where it can be seen that the pulses increase slower for proton primaries of lower energies compared 
to higher energies or electron primaries. At higher energies the time parameters for protons are very similar to those for electrons.
The reason is that protons of low energies only scintillates just after entering the plastic while more energetic protons are 
more penetrating and scintillates all along the plastic as electrons do. 

The pulse width defined as t$_{90}$-t$_{10}$ is also shown in Table \ref{table:summary} and, in this case, it is also energy 
independent for both primaries since the previous trend analyzed for protons is the same for both t$_{10}$ and t$_{90}$ and it 
is canceled. 

In the case of a scintillator without coating the only difference is that the time parameters are lower since the pulse is narrower 
as a consequence of the fact that photons tend to escape instead of reaching the detector after many reflections.

\section{Other design variables}\label{sec:othervariables}

\subsection{Surface Roughness}\label{sec:roughness}

We have modified the roughness parameter of the plastic-coating interface to check the influence in our results. If it set to 0 (completely diffuse, 
Lambertian reflection) there are no significant differences. In \cite{LoMeo} it is found that the experimental results are better reproduced with 
this parameter set to 0 (but they use crystal scintillators) which support our choice. If it is set to 1 (perfectly polished, reflection according 
to Fresnel's equations) the pulse parameters do not change significantly but the collected signal at the photo-detector increases $\sim$3\% for 
any primary type or energy. This is in agreement with results shown in \cite{Riggi} and with the usual recommendation of polishing carefully 
the plastic surface before applying the coating.

\subsection{Coating reflectivity}\label{sec:reflectivity}
 
We have check how the coating reflectivity affect the results. The simulations are performed for the intermediate plastic volume $F=3$.
Previously we have set the reflectivity to 0.90 but, as commented, it could depend on the coating thickness since TiO$_2$ fraction is constant 
in weight. As it is shown in Table \ref{table:reflectivity}, the fraction of photons that reach the photo-detector after reflections increases 
with the reflectivity as expected and, as a consequence, also does the fraction of photons detected. Since photons could be detected after more
reflections as reflectivity increases, the pulse gets wider and the time parameters increase.

\begin{table}[h!]
\centering
\caption{Varying coating reflectivity. For 0.90 values are taken from Table \ref{table:summary}.}
\label{table:reflectivity}
\begin{tabular}{|c|c|c|c|c|c|c|}
\hline
\textbf{Reflectivity} & \textbf{F$_{ref}$} & \textbf{\begin{tabular}[c]{@{}c@{}}Fraction of \\ detected  \\ photons (\%)\end{tabular}} & \textbf{t$_{10}$ (ns)} & \textbf{t$_{50}$ (ns)} & \textbf{t$_{90}$ (ns)} & \textbf{\begin{tabular}[c]{@{}c@{}}Pulse \\ witdh (ns)\end{tabular}} \\ \hline
\textbf{0.85}                 & 8.5                & 10                                                                                    & 1.8               & 4.5               & 10.5              & 8.3                       \\ \hline
\textbf{0.90}                   & 11                 & 12                                                                                    & 1.9               & 4.8               & 10.8              & 8.9                       \\ \hline
\textbf{0.93}                  & 16                 & 18                                                                                    & 2.5               & 5.5               & 12.0                & 9.2                       \\ \hline
\textbf{0.96}                  & 20                 & 22                                                                                    & 2.7               & 6.5               & 13.5              & 11.0                        \\ \hline
\end{tabular}
\end{table}

\subsection{Plastic with slow and fast decay components}\label{sec:slowfast}

The de-excitation of the plastic scintillator used in our analysis is well describe by only one decay component. However, is quite usual that
plastic scintillators emit light in two components with different decay times, usually called the fast and slow ones. Following \cite{Riggi},
we have selected the fast one to be 3.6 ns (the value used before), the slow one to be 14.2 ns and the fast/slow ratio is set to 0.73. We have 
performed a new set of simulations for the standard coating thickness (0.25 mm) and 100 MeV electrons as primaries.

\begin{figure}[h]
\centerline{
\subfigure{\includegraphics[width=8.5cm]{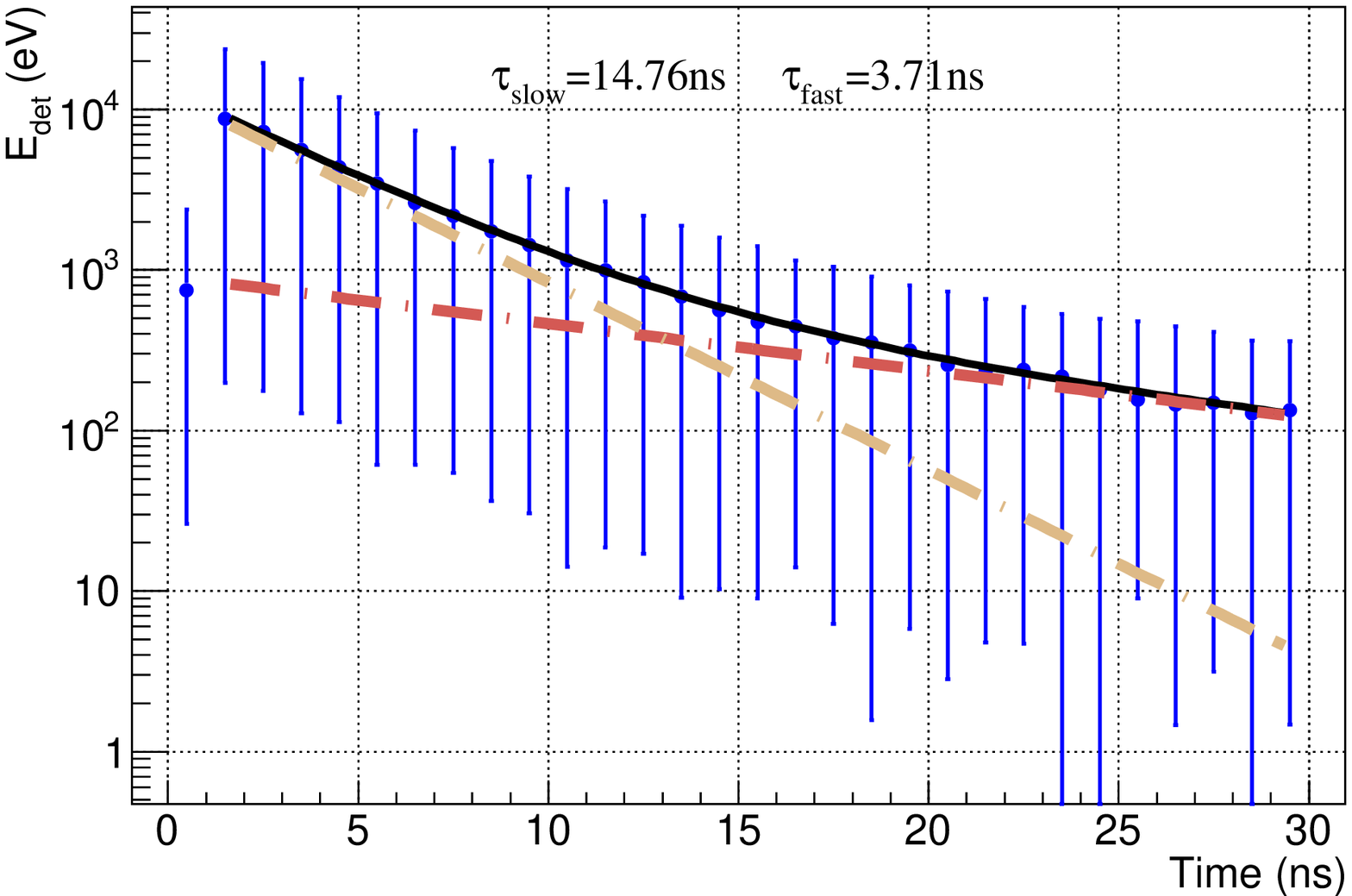}}
\subfigure{\includegraphics[width=8.5cm]{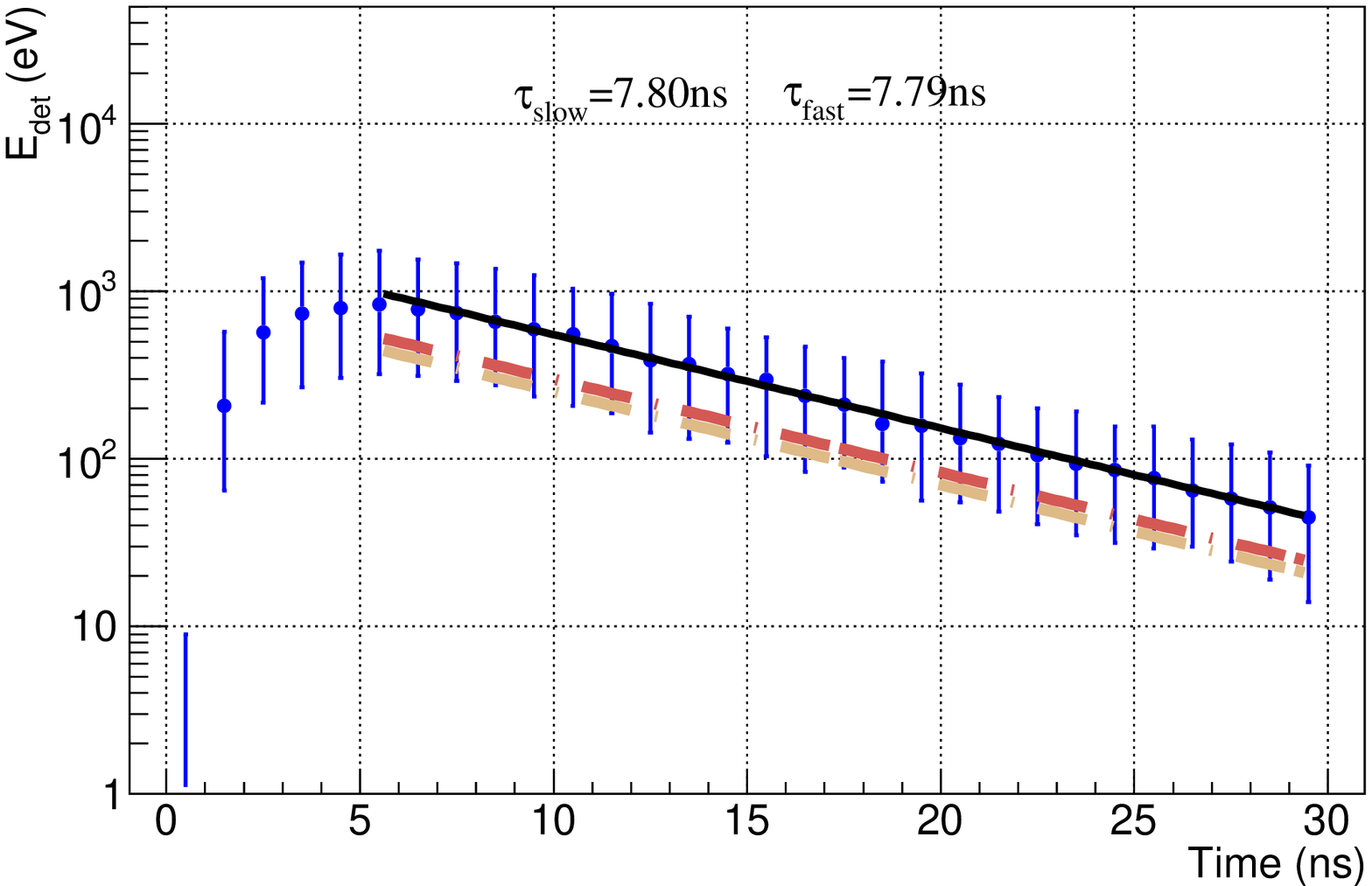}}
}
\caption{Two exponential fit to the tail of the pulse for 1 GeV protons and plastic volumes F=1 (left) and F=10 (right).}
\label{fig:TwoDecay}
\end{figure}

The pulse tail is very well fitted with two exponentials, $N_1e^{-t/\tau_{slow}}+N_2e^{-t/\tau_{fast}}$ (the only constrain to the fit is that $N_1$ 
and $N_2$ must be positive), as shown in Fig. \ref{fig:TwoDecay}. In the case of F=1 the decay parameters from the fit reproduced correctly the input 
parameters while it is not the case for larger plastics since the pulse gets wider (the same was found previously with only one decay component, 
see Fig. \ref{fig:timedecay}). The time parameters of the pulse also increases as expected as shown in Table \ref{table:twodecay}.

\begin{table}[h]
\centering
\caption{Time parameters of the pulse with two decay components (in parenthesis for one decay time, taken from Table \ref{table:summary}).}
\label{table:twodecay}
\begin{tabular}{|c|c|c|c|c|}
\hline
\textbf{F}  & \textbf{t$_{10}$ (ns)} & \textbf{t$_{50}$ (ns)} & \textbf{t$_{90}$ (ns)} & \textbf{Pulse width (ns)} \\ \hline
\textbf{1}  & 1.3 (1.2)         & 4.0 (3.3)         & 13.4 (9.0)        & 12.1 (7.8)               \\ \hline
\textbf{3}  & 2.4 (1.9)         & 6.4 (4.8)         & 15.6 (10.8)       & 18.0 (8.9)               \\ \hline
\textbf{10} & 3.0 (2.9)         & 8.1 (7.2)         & 18.7 (15.9)       & 15.7 (13.0)              \\ \hline
\end{tabular}
\end{table}

\section{Summary and Discussion}\label{sec:discussion}

\underline{Reflective coating}

The use of a reflective coating is the most common technique to increase the light collection, the key point in any of these experiments. We
have obtained that the fraction of photons lost without coating is 62, 94.5 and 99.7\% for $F$=1, 3 and 10 respectively compared to the total 
number of photons produced. As a consequence, the total signal collected at the photo-detector is reduced between 30 to 80\% (depending
on the particle type, energy and plastic volume) compared to the case in which the coating is used. It is therefore evident that it is highly advisable to 
use the coating. In addition, we have checked that if the plastic-coating interface is perfectly polished or the coating reflectivity is
enhanced, the collected light will increase (see Sec. \ref{sec:othervariables}).

A crucial issue that has not been analyzed previously in the literature is that the coating thickness imposes a lower limit to the energy 
of the primaries that could traverse it and, therefore, that is able to produce a detectable signal. This is of great importance in 
astrophysics experiments. We have determined that, with the standard thickness of 0.25 mm, 1 MeV protons do not enter the plastic as well 
as $\sim$50\% of 10 MeV protons. Protons of higher energies as well as electrons lose energy in the coating (that is not transferred to 
scintillation) in a significant amount. This effect is less important as primary energy increases. The case of protons is illustrative of
what would happen with other heavier nuclei. If coating is thicker (0.50 mm) electrons can lose up to 10\% of energy in it and only
protons with E$>$10MeV will cross it. If it is thinner (0.15 mm) electrons of 300 keV start to be detectable. Details could be found 
in Section \ref{sec:Energy}.

\underline{Plastic volume}

The comparison for the different plastic volumes is summarized in table \ref{table:summary}. Some remarks are given next:
\begin{itemize}

 \item Fraction of energy deposited in the plastic: it increases with volume but it depends strongly on primary type and energy since the 
 probability of escaping from the plastic increases with primary energy and interactions in the plastic are very different for electron 
 and proton primaries. It depends also on the energy lost in the coating as commented previously. More details in Section \ref{sec:Energy}.
 \item Technical attenuation length increases with plastic volume (Sec \ref{sec:Absorption}). This could be important when considering 
 wether an optical fiber is used to reduce the light attenuation and to increase light collection.
 \item The attenuation efficiency ($\varepsilon_{att}$) represents the fraction of photons that are not absorbed due to optical absorption. It
 decreases to 77.1\% for the largest plastic (details in Sec \ref{sec:Absorption}).
 \item The amount of light (compared to the total number of photons produced) collected directly by the photo-detector ($F_{geom}$) is 
 quite low and most of the light is collected after several reflections ($F_{ref}$). More details in  Section \ref{sec:Efficiency}.
 \item The fraction of detected photons (compared to the total number of photons produced) reduces dramatically when increasing the plastic volume
 (Sec. \ref{sec:Pulse}). The main reason is that the number of reflections grows up very rapidly with plastic volume and there exist a 10\% 
 probability of absorption in the plastic-coating interface (see discussion in Sec. \ref{sec:Surfaces}). In addition, the fraction of photons
 absorbed due to optical absorption also increases for larger plastics as commented. That leads to 40-60\% and 80-95\% less signal for $F$=3 
 and 10 compared to the smallest plastic respectively.
 \item Pulse time parameters increase with plastic volume as expected. While they are not energy dependent for electrons, they are for protons,
 except for the pulse width. More details in Sec. \ref{sec:TimeParam}. These parameters could be convoluted with the time response of the 
 photo-detector selected in each experiment to simulate the response of the entire assembly.
 
 \item The case of a plastic scintillator with two decay time components (fast/slow) has also been analyzed. The pulse tail is properly fitted
 with two exponential functions whose parameters reproduce better the input values in case of small plastic volumes. See details in 
 Sec. \ref{sec:slowfast}.
 
 \end{itemize}
 
\newpage
\begin{table}[h]
\caption{Summary: comparison for the three plastic volumes analyzed (with 0.25 mm coating)}
\vspace{0.5cm}
\begin{center}
\begin{tabular}{|c|c|c|c|} 
\hline
{\bf Volume factor ($F$)} & {\bf1} & {\bf3} & {\bf10} \\
\hline
\hline
{\bf Fraction (\%) of energy deposited} &  		&    	& 	\\
{\bf in the plastic (e-/p) (Sec. \ref{sec:Energy})}		&  		&    	& 	\\
 1 MeV   		& 85 / 0	 	& 89 / 0  & 89 / 0 	\\
 10 MeV   		& 25 / 46		& 66 / 46 & 86 / 47 \\
 30 MeV   		& 7 / 88		& 25 / 88 & 64 / 88 \\
 100 MeV  		& 2 / 20 		& 10 / 75 & 23 / 99 \\
 1 GeV   		& 0.2 / 0.8	& 0.3 / 2 & 2 / 5 	\\   
\hline
\hline
{\bf Optical Absorption  (Sec. \ref{sec:Absorption})}	&  	&    & 	 \\
Technical Absorption  	&  	&    & 	\\
 Length (cm)   		&  3.5	& 28.0   & 81.8	 \\
 $\varepsilon_{att}$ (\%)  	& 98.5 	& 88.7    & 77.1 	 \\
\hline
\hline
{\bf Light collection factors  (Sec. \ref{sec:Efficiency})}&  	&    & 	\\
 $F_{geom}$ (\%)  & 8 	&  0.8  & $\leq$0.05	 \\
 $F_{ref}$ (\%)  &  50	& 11    & 0.5	 \\
 \hline
 \hline
 {\bf Light at the photo-detector  (Sec. \ref{sec:Pulse})} &  	&    & 	 \\
 Fraction (\%) of detected photons	&  70		&  12  	& 0.8	\\
Integrated signal (normalized to $F$=1)	&  	1	& 0.4-0.6$^{*}$   	& 0.05-0.20$^{*}$	\\
\hline
\hline
{\bf Pulse parameters (Sec. \ref{sec:TimeParam})}&  	&    & 	\\
 Decay time (plastic 3.6 ns) 	& 3.6  		&  3.7-3.8  	& 5.6-5.9	\\	
 $t_{10}$ (ns)$^{**}$ 		& 1.2  		&  1.9  	& 2.9	\\	
 $t_{50}$ (ns)$^{**}$	& 3.3		&  4.8   	& 7.2	\\
 $t_{90}$ (ns)$^{**}$		&9.0		&  10.8  	& 15.9	\\
 $t_{90}-t_{10}$	&7.8		&8.9		&13.0	\\
\hline
\end{tabular}
\end{center}
\label{table:summary}
\begin{tablenotes}
 \item[] * Vary with primary energy
 \item[] ** For protons vary with energy (see Sec. \ref{sec:TimeParam}).
\end{tablenotes}
\vspace{0.5cm}
\end{table}

\newpage
\section{Conclusions}\label{sec:conclusions}

We have studied how the plastic volume and the use of a reflective coating will affect the capabilities of an experiment based on the use of 
plastic scintillators in the context of an astrophysics experiments, though the analysis is also useful for medicine or particle physics where
plastic scintillators are traditionally used. Previous studies are based on experimental setups with different plastics, 
geometries, beams or photo-detectors. On the other hand, we focus on a more general approach using the highly tested GEANT4 simulation tool 
which allows to easily consider a full set of material properties and geometries.

The best choice will depend on the scientific objectives of the experiment depending on the primary type and energy. The thickness of the 
coating imposes a lower limit to the primary energy that would be detectable that also depends on primary type. In general, if the plastic volume
is comparable to the size of the photo-detector in contact with it, the collected signal will be enhanced but particle type or energy discrimination 
will be more difficult and a larger exposure time could be needed. In case of astroparticle physics experiments in orbit, the use of smaller plastics 
is a great advantage to reduce the costs. 

Several parameters have been analyzed such as the energy lost in the coating, the deposited energy in the scintillator, the optical absorption, 
the fraction of scintillation photons that are not detected, the light collection at the photo-detector, the pulse shape and its time parameters 
and finally, other design parameters as the surface roughness, the coating reflectivity and the case of a scintillator with two decay components.
This work could serve as a guide on the design of future experiments based on the use of plastic scintillators.

\section{Acknowledgments}

GMT thanks the support of CONACyT under grant CB-239660 and CB-207065 and Universidad Nacional Aut\'onoma de M\'exico under grant PAPIIT-IN103217.
ADS is member of the Carrera del Investigador Cient\'ifico of CONICET, Argentina.

\end{document}